\documentclass[pre,preprint,superscriptaddress,showpacs,endfloats*]{revtex4}

\usepackage[]{graphicx} 
\usepackage{subfigure}

\def\be{\begin{equation}}
\def\ee{\end{equation}}
\def\bea{\begin{eqnarray}}
\def\eea{\end{eqnarray}}
\def\II{\hbox{$1\hskip -1.2pt\vrule depth 0pt height 1.6ex width 0.7pt\vrule depth 0pt height 0.3pt width 0.12em$}}

\newcommand{\acrits}{\alpha_\mathrm{crit}}
\newcommand{\fco}{f_{c}}
\newcommand{\reffig}[1]{\mbox{Fig.~\ref{#1}}}
\newcommand{\refeq}[1]{\mbox{Eq.~(\ref{#1})}}
\newcommand{\refsec}[1]{\mbox{Sec.~\ref{#1}}}
\newcommand{\reftab}[1]{\mbox{Tab.~\ref{#1}}}

\begin{document}

\hyphenation{re-so-nan-ce re-so-nan-ces ex-ci-ta-tion z-ex-ci-ta-tion di-elec-tric ap-pro-xi-ma-tion ra-dia-tion Me-cha-nics quan-tum pro-posed Con-cepts pro-duct Reh-feld ob-ser-va-ble Se-ve-ral rea-so-nable Ap-pa-rent-ly re-pe-ti-tions re-la-tive quan-tum su-per-con-duc-ting ap-pro-xi-mate cri-ti-cal func-tion wave-guide wave-guides}

\title{Bound States in Sharply Bent Waveguides: Analytical and Experimental Approach}
\author{S. Bittner}
\author{B. Dietz}
\email{dietz@ikp.tu-darmstadt.de}
\author{M. Miski-Oglu}
\affiliation{Institut f\"ur Kernphysik, Technische Universit\"at Darmstadt, D-64289 Darmstadt, Germany}

\author{A. Richter}
\email{richter@ikp.tu-darmstadt.de}
\affiliation{Institut f\"ur Kernphysik, Technische Universit\"at Darmstadt, D-64289 Darmstadt, Germany}
\affiliation{ECT*, Villa Tambosi, I-38123 Villazzano (Trento), Italy}

\author{C. Ripp}
\affiliation{Institut f\"ur Kernphysik, Technische Universit\"at Darmstadt, D-64289 Darmstadt, Germany}

\author{E. Sadurn\'\i}
\affiliation{Instituto de F\'isica, Benem\'erita Universidad Aut\'onoma de Puebla, C. P. 72570 Puebla, M\'exico}
\affiliation{Institut f\"ur Quantenphysik and Center for Integrated Quantum Science and Technology, Universit\"at Ulm, Albert-Einstein-Allee 11, D-89081 Ulm, Germany}

\author{W. P. Schleich}
\affiliation{Institut f\"ur Quantenphysik and Center for Integrated Quantum Science and Technology, Universit\"at Ulm, Albert-Einstein-Allee 11, D-89081 Ulm, Germany}
\date{\today}

\begin{abstract}

Quantum wires and electromagnetic waveguides possess common features since their physics is described by the same wave equation. We exploit this analogy to investigate experimentally with microwave waveguides and theoretically with the help of an effective potential approach the occurrence of bound states in sharply bent quantum wires. In particular, we compute the bound states, study the features of the transition from a bound to an unbound state caused by the variation of the bending angle and determine the critical bending angles at which such a transition takes place. The predictions are confirmed by calculations based on a conventional numerical method as well as experimental measurements of the spectra and electric field intensity distributions of electromagnetic waveguides.
\end{abstract}



\maketitle

\section{Introduction}

The opinion that a binding potential is necessary to obtain a bound state of a particle in quantum mechanics~\cite{Bohm1951} was already exposed as being incomplete in the early days of quantum mechanics. Indeed, in 1929 John von Neumann and Eugene P. Wigner~\cite{Neumann1929} showed that even a strongly repulsive potential can support a bound state with a normalizable wave function. Since then many examples of unusual bound states in various quantum and classical analogue systems with different bindung mechanisms have been investigated. For an overview see for example Ref.~\cite{Cirone2001a}.  

In quantum mechanics it is by now well-known that a binding potential in two space dimensions always supports a bound state no matter how shallow the potential, whereas in three space dimensions the potential needs to have a certain depth. The reason is that in two space dimensions the state of vanishing angular momentum, that is, a rotationally symmetric state, brings in an additional attractive potential which decays quadratically with the inverse of the distance. This anti-centrifugal potential~\cite{Cirone2001} arises from the adaption of the Laplacian to the symmetry of the state and appears in many other situations where the boundary conditions or the preparation mechanism dictate the symmetry of the state. In the present paper we develop an approximate but analytical technique to describe the occurrence of such geometry-induced bound states for the case of a bent waveguide. We compare and contrast our analysis with numerical calculations based on a conventional method as well as with measurements of these bound states.  

\subsection{From compound nucleus to bent waveguides}
\begin{figure}[bt]
\begin{center}
\subfigure[]{
	\includegraphics[width = 6 cm]{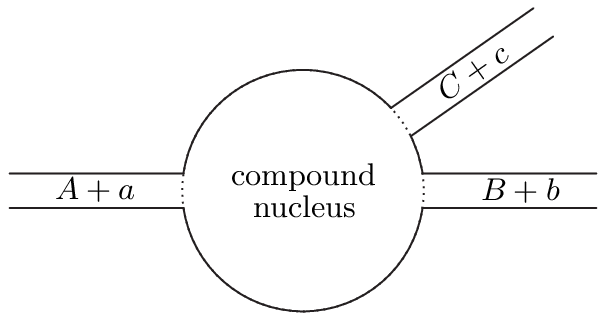}
	\label{sfig:schemCompNucl}
}
\subfigure[]{
	\includegraphics[width = 6 cm]{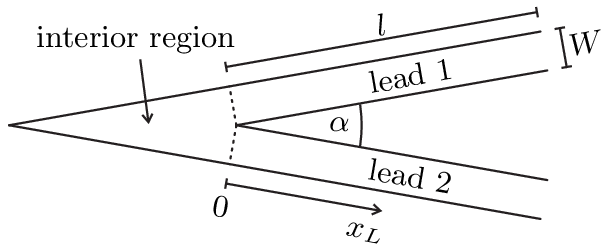}
	\label{sfig:schemWG}
}
\end{center}
\caption{\label{fig:schematics} Analogy between the decay of a compound nucleus \subref{sfig:schemCompNucl} and a bound state in a bent waveguide~\subref{sfig:schemWG}. A compound nucleus is formed by bombarding a nucleus $A$ with a particle $a$ and eventually decays into a residual nucleus and particles, $B+b$, $C+c$, etc.. This coupling to the exterior region is modeled by attaching different formation and decay channels to a confined, interior region. Panel \subref{sfig:schemWG} shows a sketch of a quantum wire or electromagnetic waveguide with a sharp bend of angle $\alpha$, the straight parts of which have a length $l$ and a width $W$. The border between them and the corner region is indicated by dashed lines. Since bound states form in the latter, we also denote it as interaction region. The coordinate along the leads is denoted by $x_L$.}
\end{figure}

Systems of the type considered in the present paper in fact serve as models for the understanding of properties of the scattering processes associated with compound nucleus reactions. Within such a scattering approach the compound nucleus is modeled by a confined region of quasibound states, and a number of open reaction channels attached to it. These represent the coupling of the quasibound states to the continuum, i.e., their formation and decay~\cite{Kapur1938, Feshbach1958, Lane1958, Vogt1962, Mahaux1969, Mitchell2010}~[see \reffig{sfig:schemCompNucl}]. 

The universal scattering properties of a compound nucleus reaction, in turn, can be simulated in microwave experiments, where electromagnetic waves propagate via antennas or waveguides into a cavity corresponding to the interior region which allows for bound or quasibound states~\cite{StoeckmannBuch2000, Beck2003, Mitchell2010, Dietz2010}. The most elementary example is a bent waveguide with a corner as depicted in \reffig{sfig:schemWG} --- a model for strong interactions in nuclei similar to that one was proposed in Ref.~\cite{Lenz1986}. States trapped around the inner corner were predicted to exist below the energy of the first propagating mode~\cite{Lenz1986, Schult1989, Exner1989, Exner1989b, Goldstone1992} and observed experimentally in bent microwave waveguides~\cite{Carini1992, Carini1993, Carini1997a} and in quantum wires~\cite{Wu1991,Wu1993,Wang1995}. Note that the Helmholtz equation describing the physics of the former and the Schr\"odinger equation which is applicable to the latter are mathematically identical in the energy range of the bound states. This analogy was used in the numerical computation of bound states in bent waveguides~\cite{Carini1993} and quantum wires~\cite{Weisshaar1989,Weisshaar1991,Wang1995}. The experimental and the numerical studies revealed that with the variation of the waveguide's bending angle $\alpha$ [see Fig.~\ref{sfig:schemWG}] a transition of the modes from bound to propagating ones takes place. The bending angles at which such a transition occurs are called critical angles in the following. 

\subsection{Formulation of the problem}
In this paper we present a detailed theoretical and experimental study of that transition. To test the analytical approach, existing results~\cite{Carini1993} were extended down to bending angles $\alpha =0.2^\circ$ using an improved conventional numerical method which is based on a scattering approach. Its main achievement is an analytical formula for the number of bound states at a given bending angle derived from a fit to the numerical result. The experiments performed to thoroughly test these results comprised measurements of the eigenfrequencies (energies) and the associated field intensity distributions (squared wave functions) of the bound states for several bending angles. Furthermore, we present experimental results concerning the effect of the finite lead lengths on the bound states and the occurrence of resonant tunneling. These investigations go beyond the numerical and theoretical ones which assume waveguides of infinite lengths. Only recently the existence of trapped modes in finite waveguides, as used in the experiments, was studied theoretically~\cite{Delitsyn2012}. 

The numerical computations of the bound states and the critical angles of a waveguide or quantum wire involve the solution of the Schr\"odinger equation of a free particle with Dirichlet boundary conditions imposed on its wave functions at the walls of the system~\cite{Schult1989,Exner1989a,Exner1989b,Weisshaar1989,Weisshaar1991,Goldstone1992,Carini1993,Amore2012}. 

The theoretical treatment of a problem of this kind presented in Ref.~\cite{Sadurni2010} encompasses a conformal mapping applied to the geometry of the bent waveguide which induces in the Schr\"odinger equation an additional effective potential given by the associated Jacobian. Starting from this equation the bound states at a given bending angle and the critical angles were determined using a Wentzel-Kramers-Brillouin (WKB) approximation developed in Ref.~\cite{Bestle1995} for systems containing sharp corners. The results agree well with those obtained experimentally with electromagnetic waveguides and from the numerical computations based on a scattering approach.  

\subsection{Relation to other work}
It is worthwhile to mention again that some of the salient features of the spectrum of a bent waveguide were already understood in previous experiments with electromagnetic waveguides~\cite{Carini1992, Carini1993, Carini1997a} and quantum wires~\cite{Wu1991,Wu1993}. Nevertheless, we feel that it is important to obtain a complete solution, by improving the conventional numerical and experimental methods, which encompasses all the important features and providing a general method which does not rely on numerical calculations. For example, one may pose the following question: What is the number of bound states supported by an orthogonal cross-wire configuration? The numerical answer is two~\cite{Schult1989}, but the reason for this should be revealed by a suitable treatment relating the shape of a wire to its binding capabilities. 

These effects, rather than being just accidental, are linked to deeper concepts in undulatory physics. The confinement of a particle to a restricted domain by introducing walls is directly related to a proper formulation of the d'Alembert principle in quantum mechanics~\cite{Koppe1971,Jensen1971}. This problem must be carefully tackled in the light of well-known results such as the trapping of waves in smoothly bent quantum wires~\cite{Exner1989, Exner1989a, Goldstone1992, Bulgakov2002} and the emergence of a so-called curvature-induced quantum potential~\cite{daCosta1981}. For instance, it is striking that in bent waveguides or quantum wires with open ends as depicted schematically in \reffig{sfig:schemWG}, semiclassical methods such as trace formulas derived for closed~\cite{Gutzwiller1990,Bogomolny1992} and open systems~\cite{Miller1974,Doron1991,Schwieters1996} fail because there are no closed orbits and there is no classical counterpart including a binding force. We present an analytical approach which yields a Hamiltonian whose spectrum comprises the bound states. The conclusion to be drawn from this analysis is that the mere presence of curved boundaries may induce a binding potential, with significant implications close to sharply bent boundaries. 

\subsection{Outline of the article}
The article is organized as follows: in \refsec{sec:num} we shortly review the results of our calculations of the bound states in a bent waveguide and of the critical angles based on a conventional method. Section~\ref{sec:exp} comprises the description of the microwave experiments and the comparison of the experimental and the numerical results from~\refsec{sec:num}. Section~\ref{sec:theo} presents analytical estimates of the critical angles and the number of bound states in a bent waveguide. The experimental and the numerical results presented in Secs.~\ref{sec:num} and~\ref{sec:exp} demonstrate that for a given bending angle the wave functions of the bound states with excitation frequency close to the cut-off frequency penetrate considerably into the leads. Similarly, the influence of the binding potential revealed in the theoretical approach of~\refsec{sec:theo} reaches into the leads. In~\refsec{sec:expAdd} this phenomenon is investigated experimentally. The conclusions are summarized in~\refsec{sec:concl}.

\section{\label{sec:num}Numerical Results}

In this section we compute the eigenenergies and wave functions for the bound states of sharply bent quantum wires, that is, the eigenfrequencies and the associated electric field strengths for those of sharply bent rectangular waveguides using a conventional method. For this we use the fact that below the frequency $f_{\rm max}=c/(2h)$, with $c$ the speed of light and $h$ the height of the waveguide, only the TE$_{1,0}$-mode is excited~\cite{Note1}. There, the Helmholtz equation associated with a bent waveguide is mathematically identical to the stationary Schr\"odinger equation of an open quantum billiard or a quantum wire of corresponding shape~\cite{StoeckmannBuch2000,Richter1999,Sridhar1992}. The eigenfrequencies of the bound states are all below the cut-off frequency $f_c=c/(2W)$ of the first propagating mode, where $W$ is the width of the waveguide. Consequently, since rectangular waveguides with the width larger than the height were used in the experiments, such that $f_{\rm max}>f_c$, the numerical solutions of the stationary Schr\"odinger equation obtained in this section apply to bent quantum wires~\cite{Weisshaar1989,Weisshaar1991,Wu1993,Wang1995} and also to bent microwave waveguides.

The stationary Schr\"odinger equation of an open quantum billiard with the shape of a bent waveguide corresponds to that for a free particle with Dirichlet boundary conditions along the boundary $\partial\Omega$ defined by the walls of the waveguide in \reffig{sfig:schemWG}. Here, $\Omega$ denotes the interior of the wave\-guide. A coordinate system $(x, y)$ is introduced in the plane of the waveguide with origin at the outer corner and the $x$-axis along the symmetry axis of the waveguide to write for $x,\, y\in\Omega$ 
\be \label{eq1} -\Delta_{x, y}\phi(x,y)=E\phi(x,y), \qquad \phi|_{\partial\Omega} = 0, \ee
where we use units $\hbar^2/(2m)=1$, the energy $E$ corresponds to the square of the wave number $k$, $E=k^2$, the wave function $\phi (x,y)$ denotes the $z$-component of the electric field vector in the waveguide and 
\be
\Delta_{x,y}=\frac{\partial^2}{\partial x^2}+\frac{\partial^2}{\partial y^2} 
\ee
is the two-dimensional Laplacian. The width of the waveguide is set to $W=1$ in all calculations. 

\begin{figure*}[bt]
\begin{center}
\subfigure[]{
        \includegraphics[width = 8.4 cm]{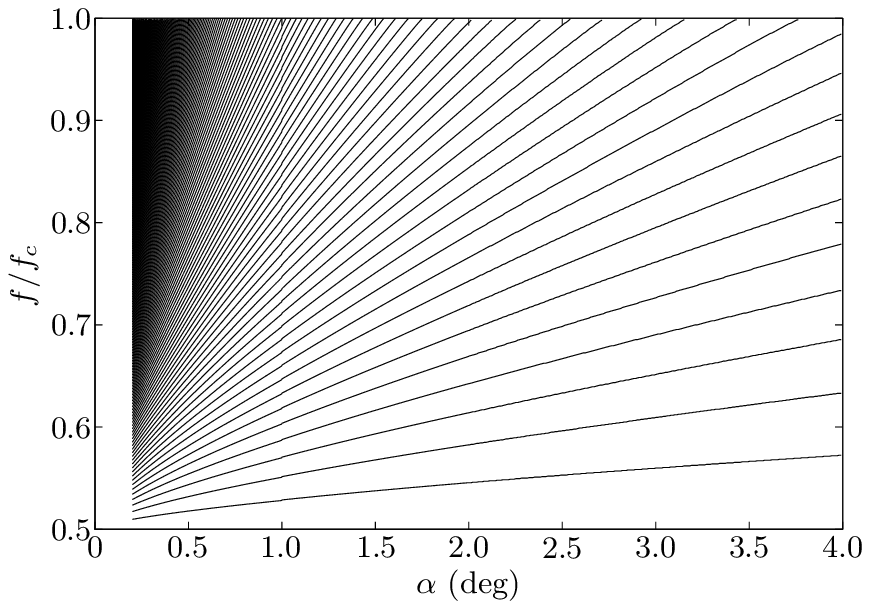}
        \label{sfig:Num1a}
}
\subfigure[]{
        \includegraphics[width = 8.4 cm]{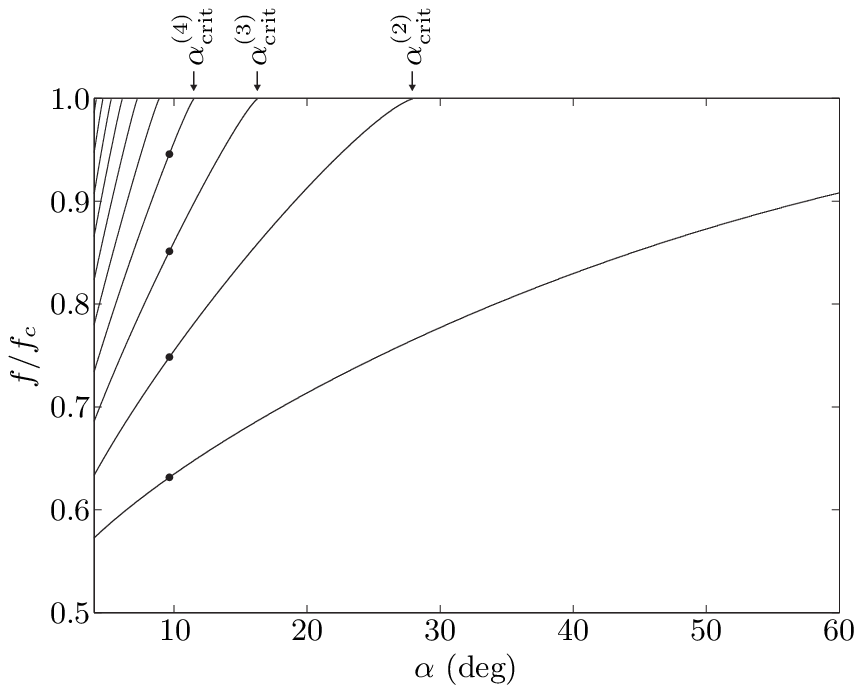}
        \label{sfig:Num1b}
}
\end{center}
\caption{\label{Num1} Numerical results for the rescaled eigenfrequencies $f/f_c$ with $f$ the eigenfrequencies of the bound states and $f_c$ the cut-off frequency versus the bending angle $\alpha$ of the waveguide. Figure~\ref{sfig:Num1a} shows the ratios for $0.2^\circ\leq\alpha\leq 4^\circ$, Fig.~\ref{sfig:Num1b} those for $4^\circ\leq\alpha\leq 60^\circ$. Each curve corresponds to the eigenfrequency of one bound state in units of $f_c$, which is a continuous function of the bending angle $\alpha$, starts at $\alpha =0^\circ$ and increases until it reaches the value $f/f_c=1$ at a certain bending angle. There the bound state turns into a propagating mode. The corresponding angles are called critical angles $\acrits$. The arrows in panel (b) indicate the critical angles considered in the experiments with microwave waveguides. The number of bound states increases rapidly as $\alpha$ approaches $0^\circ$. Indeed, for $\alpha\approx 0.2^\circ$ the curves are barely distinguishable. The eigenfrequency of the first bound state, which corresponds to the lowest curve in both panels, converges to $f/f_c=1/2$. This is the cut-off frequency of a waveguide of double the width of the leads. The dots mark the eigenfrequencies of the four bound states at $\alpha=9.65^\circ$, whose intensity distributions are shown in Fig.~\ref{Num3}.}
\end{figure*}

The procedure for the numerical determination of the eigenenergies and eigenfunctions of \refeq{eq1} is outlined in Appendix~\ref{ssec:numerics}. It is based on the scattering approach described in Refs.~\cite{Weisshaar1989,Weisshaar1991,Doron1991,Dietz1992}. For this purpose the waveguide is subdivided into an interior region and an asymptotic region defined by its straight leads with parallel walls as indicated in Fig.~\ref{sfig:schemWG}. The procedure is similar to that used for the description of compound nucleus reactions in the framework of scattering theory, where the interior region corresponds to the compound nucleus and the leads to the formation and decay channels~\cite{Mahaux1969, Dittes2000}. 

Figure~\ref{Num1} shows the numerical values $k/k_c=f/f_c$ for the bound states as function of the bending angle $\alpha$. Here, $k = \sqrt{E}$ is the wave number, and $f=ck/(2\pi)$ with $c$ the speed of light is the corresponding frequency. At the cut-off frequency $f_c=ck_c/(2\pi)= c/(2W)$ the first propagating mode emerges in the leads [see \refeq{eq:4}]. Each curve corresponds to the eigenfrequency of one bound state which varies continuously with the bending angle $\alpha$. The number of bound states at a bending angle $\alpha$ is obtained by determining in Fig.~\ref{Num1} the number of curves at $\alpha$. All curves start at $\alpha =0^\circ$ and increase until they eventually reach the value $f/f_c=1$ at a critical angle $\acrits$. There, the number of bound states decreases by one. Above $\acrits^{(2)}\approx 28.28^\circ$, with $\acrits^{(n)}$ denoting the critical angle of the $n$th bound state, only one bound state survives. It corresponds to the lowest curve in Fig.~\ref{Num1} and has $\acrits^{(1)}=180^\circ$. With decreasing $\alpha$ the number of bound states at a given value of $\alpha$ increases rapidly. Indeed, as can be observed in Fig.~\ref{sfig:Num1a}, it is hardly possible to resolve all curves around $\alpha \approx 0.2^\circ$ due to their large number. For $\alpha\to 0^\circ$ the eigenfrequency of the lowest bound state approaches from above the value $f / f_c =1 / 2$, so all eigenfrequencies obey the inequality $f/f_c\geq 1/2$. Note that the frequency $f = f_c/2$ corresponds to the cut-off frequency of a straight waveguide with twice the width of the leads. Indeed, such a waveguide is obtained by decreasing the angle of $\alpha$ to $0^\circ$ in~\reffig{sfig:schemWG}.


\begin{figure}[bt]
\begin{center}
\includegraphics[width = 8.4 cm]{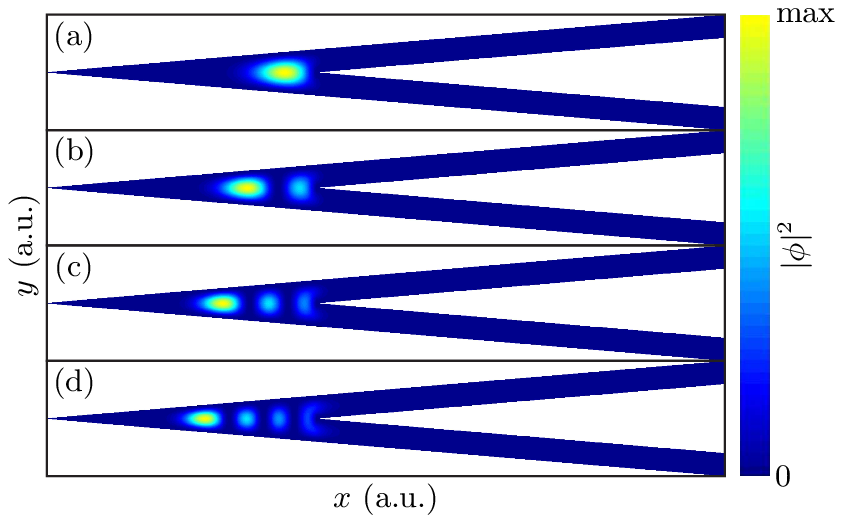}
\end{center}
\caption{\label{Num3} (Color online) Square of the numerically computed wave functions associated with the four bound states at $\alpha =9.65^\circ$. The corresponding eigenfrequencies are marked by dots in Fig.~\ref{sfig:Num1b}. Panel (a) shows the intensity distribution for $f/f_c=0.631$, panel (b) that for $f/f_c=0.748$, panel (c) that for $f/f_c=0.851$ and panel (d) that for $f/f_c=0.946$. The blue (darkest) color corresponds to the lowest, the yellow (brightest) to the highest intensity.}
\end{figure}
Figure~\ref{Num3} shows the intensity distributions corresponding to the four bound states existing at $\alpha =9.65^\circ$. These are marked by dots in Fig.~\ref{sfig:Num1b}. For the first bound state at $f/f_c=0.631$ the intensity distribution shown in panel (a) exhibits one maximum close to the inner corner. With increasing $f/f_c$, that is, for the second [panel (b)], the third [panel (c)] and the fourth [panel (d)] bound state the number of maxima increases by, respectively, one. The maximum with the highest intensity moves away from the inner corner to the outer one, and that closest to the inner corner penetrates more and more into the leads. This leakage has been investigated experimentally. The results are presented in Sec.~\ref{sec:expAdd}. 

With the numerical method presented in Appendix~\ref{ssec:numerics} we were able to compute the eigenfrequencies of the bound states of bent waveguides with bending angles as small as $0.2^\circ$. For smaller angles the numerical errors are larger than the spacing between the eigenfrequencies with values $f\simeq f_c$.

\begin{figure}
\begin{center}
\includegraphics[width = 8.4 cm]{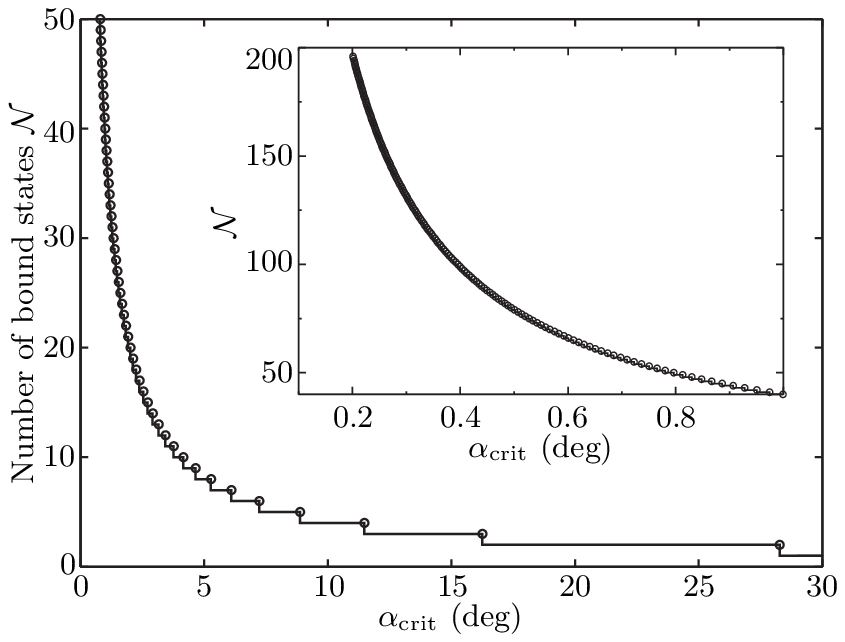}
\end{center}
\caption{\label{Num4} Number of bound states $\mathcal{N}(\alpha)$ in its dependence on the bending angle $\alpha$. The circles show the numerical results, the solid line is obtained from the analytical expression~\refeq{fitnumber}. The agreement between the numerical results and the analytical expression is very good. The inset shows a magnification for $\acrits\le 1^\circ$.}
\end{figure}

In Fig.~\ref{Num4} the number of bound states $\mathcal{N}(\alpha)$ is plotted versus the bending angle $\alpha$. The circles correspond to the numerical results. They are well described by the analytical equation 
\be\mathcal{N}(\alpha)={\rm Int}\left[\frac{a_0}{3}\csc\left(\frac{\alpha}{2}\right)+1/2\right]\label{fitnumber}\ee
with the fit parameter $a_0=1.027$ (solid line) close to unity and Int$[x]$ denoting the integer part 
of $x$.

\section{\label{sec:exp}Measurement of bound states}

In the preceding section we have presented numerically determined eigenfrequencies and associated wave functions, i.e., electric field strengths, of the bound states in bent rectangular waveguides as function of the bending angle and determined the critical angles, where a bound states turns into a propagating mode. These results were tested experimentally with bent microwave waveguides as described in the present section.   

\subsection{Experimental setup}
Experiments with microwave cavities are widely used to investigate universal properties of quantum systems such as the spectral statistics of chaotic quantum billiards \cite{Graf1992}, chaotic scattering \cite{Kuhl2005a, Mitchell2010} or fidelity decay \cite{Schaefer2005}. This approach is possible due to the analogy noted already above between the stationary Schr\"odinger equation and the Helmholtz equation for flat cylindrical microwave cavities \cite{Richter1999, StoeckmannBuch2000}. Similarly, the properties of quantum wires have been investigated in experiments with microwave waveguides of rectangular cross section \cite{Carini1992, Carini1993, Carini1997a}. Here we report on microwave experiments performed to validate the numerical and the theoretical results concerning the critical angles and the dependence of the bound states on the bending angle. 

\begin{figure}[bt]
\begin{center}
\includegraphics[width = 6 cm]{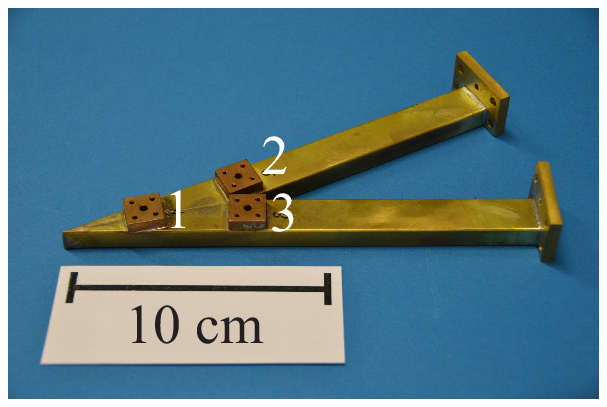}
\end{center}
\caption{(Color online) Photo of a bent waveguide with angle $\alpha = 26.50^\circ$. The three small copper blocks soldered on top of the waveguide are used to attach the antenna ports (not shown) labeled by numbers $1$, $2$ and $3$, where microwave power is coupled into and out of the waveguide. Additional straight waveguide parts can be attached to the two flanges.}
\label{fig:expSetup}
\end{figure}

Nine different waveguides with different bending angles $\alpha$ were investigated experimentally. Figure~\ref{sfig:schemWG} shows the geometry of the bent waveguides. They were constructed by soldering together two tapered WG18 brass waveguides (by Flann Microwave) as shown in \reffig{fig:expSetup}. The leads with lengths $l_0$ were extended to their full lengths $l$ by attaching additional straight waveguides with open ends. The angles of three of the nine waveguides were chosen close to a critical angle, $\acrits^{(2)}\approx 28.28^\circ$, $\acrits^{(3)}\approx 16.26^\circ$ and $\acrits^{(4)}\approx 11.48^\circ$, which were read off from~\reffig{sfig:Num1b} and are marked in this figure by arrows. The lengths $l$ were chosen larger than $400$ mm, in order to minimize the effect of the finite length of the waveguides analyzed in more detail in~\refsec{sec:expAdd}. The angles and the lengths of the different waveguides are listed in \reftab{tab:angles}. They were determined with a precision $\Delta \alpha = 0.1^\circ$ and $\Delta l = 0.5$ mm, respectively. The waveguides had a rectangular cross section with a width $W = (15.76 \pm 0.02)$ mm and a height $h = (7.9 \pm 0.02)$ mm. 

According to \refeq{eq:4}, a straight waveguide of this width allows for propagating modes above the cut-off frequency $\fco =(9.511 \pm 0.012)$~GHz. Below $\fco$, only evanescent fields can exist. Note that the quantities of interest, the ratios $f/f_c$, do not depend on the width $W$ of the waveguide. Hence, the experimental ratios may be compared with the computed ones of \refsec{sec:num} for all choices of $W$. 

Below $f_\mathrm{max} = c / (2 h) \approx 19$~GHz the electric field is perpendicular to the  waveguide plane defined by the $x$ and $y$ coordinates and described by the scalar Helmholtz equation~\cite{Jackson1999,Sridhar1992}
\be \label{eq:helmholtzEM} (\Delta_{x, y} + k^2) E_z(x, y) = 0, \qquad E_z|_{\partial \Omega} = 0\, . \ee
Equation~(\ref{eq:helmholtzEM}) is identical to \refeq{eq1} with $\phi=E_z$ and $E=k^2=\left(2\pi f/c\right)^2$. 

\begin{table}[tb]
\caption{List of the bending angles $\alpha$ of the waveguides investigated experimentally. The lengths $l$ and $l_0$ of the leads are with and without an additional waveguide extension, respectively.}
\label{tab:angles}
\begin{tabular}{c|c|c}
\hline
\hline
$\alpha$ (deg) & $l_0$ (mm) & $l$ (mm) \\
\hline
28.40 & 122.2 & 422.2 \\
27.65 & 356.7 & 456.7 \\
26.50 & 131.6 & 431.6 \\
16.40 & 211.4 & 411.4 \\
15.95 & 224.8 & 424.8 \\
14.55 & 240.0 & 440.0 \\
11.83 & 300.1 & 400.1 \\
11.00 & 328.2 & 428.2 \\
9.65  & 361.2 & 461.2 \\
\hline
\hline
\end{tabular}
\end{table}

\begin{figure}[tb]
\begin{center}
\includegraphics[width = 8.4 cm]{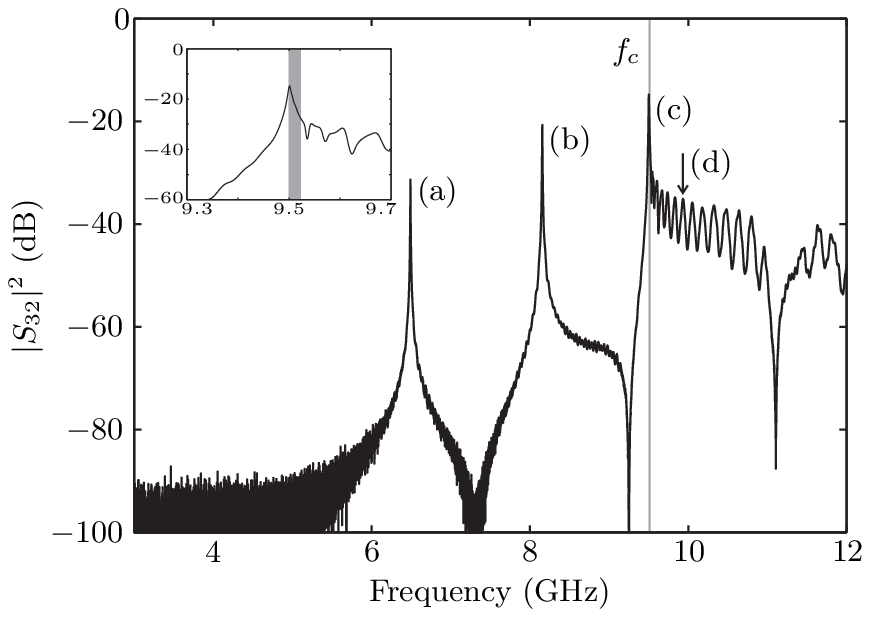}
\end{center}
\caption{Bound states of a bent waveguide, here with bending angle $\alpha = 16.40^\circ$, identified by the measured resonance spectrum shown in logarithmic scale. The gray bar indicates the frequency range where the cut-off frequency $\fco$ is expected. The broad resonances observed above $\fco$ are due to partial reflections of the microwaves at the open ends of the waveguide. The intensity distributions for the resonances labeled (a)--(d) are shown in the corresponding panels of \reffig{fig:WFexamples}. The arrow indicates the position of the resonance (d), which is above the cut-off frequency $\fco$ and thus broad. The inset is a magnification around $\fco$ including resonance (c).}
\label{fig:freqSpec}
\end{figure}

Microwave power was coupled into and out of the waveguides with two small wire antennas aligned perpendicularly to the waveguide plane. They protruded about $1.5$ mm into the waveguide through holes with a diameter of $3$~mm in its top wall at three different positions, one in the interior region (label $1$ in Fig.~\ref{fig:expSetup}) and two close to it in the leads (labels $2$ and $3$ in Fig.~\ref{fig:expSetup}). Results are presented for the measurements with antennas at positions $2$ and $3$. A vectorial network analyzer (VNA, model PNA N5230A by Agilent Technologies) was used to measure the transmission amplitude $S_{ba}(f)$ from antenna $a$ to antenna $b$, where
\be |S_{ba}(f)|^2 = \frac{P_\mathrm{out,\,b}}{P_\mathrm{in,\,a}} \ee
is the ratio of the power $P_\mathrm{out,\,b}$ coupled out via antenna $b$ and the power $P_\mathrm{in,\,a}$ coupled in via antenna $a$ at the excitation frequency $f$. 

\subsection{Resonance spectra and frequencies and field intensities}
The measured resonance spectrum of the waveguide with bending angle $\alpha = 16.40^\circ$ is shown in \reffig{fig:freqSpec}. There are three sharp resonances labeled by (a)--(c) below the cut-off frequency (gray bar) at frequencies $6.49$, $8.16$ and $9.50$~GHz. Their quality factors $Q \approx 2000$ are comparable to those of resonances in closed normal conducting microwave cavities of similar size, which indicates that these resonances correspond to the three bound states predicted for this waveguide. Above $\fco$ the spectrum exhibits a series of broad resonances like the one labeled by (d). These are due to partial reflections of the microwaves at the open ends of the waveguide.


\begin{figure}[tb]
\begin{center}
\includegraphics[width = 8.4 cm]{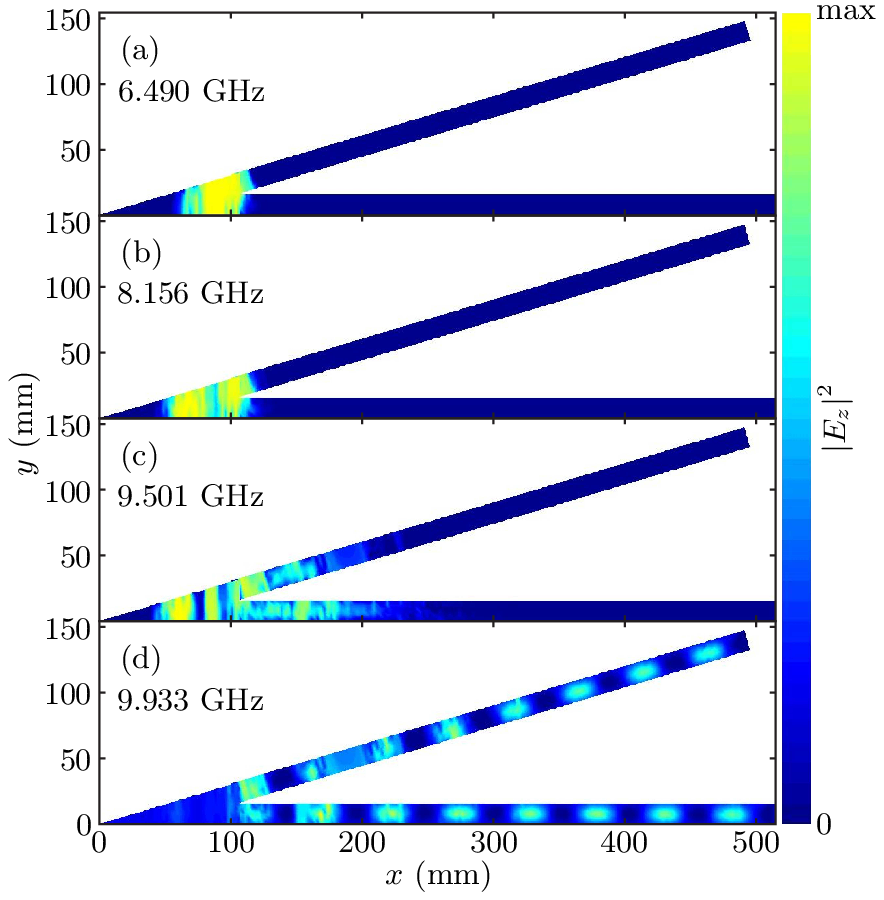}
\end{center}
\caption{(Color online) Intensity distributions of three bound states and one propagating mode for the waveguide with $\alpha = 16.40^\circ$ and length $l=411.4$~mm measured with the perturbation body method. The electric field intensity $I\propto\vert E_z\vert^2$ is plotted in false colors. The blue (darkest) color corresponds to the lowest, the yellow (brightest) to the highest intensity. Panels (a)--(d) correspond to the resonances indicated in \reffig{fig:freqSpec}. In panels (a)--(c), i.e., for $f<f_c$, the intensity distribution is localized around the inner corner, whereas for $f>f_c$ it periodically extends over the whole waveguide as shown for one example in panel (d). In panels (a) to (c) yellow (bright) stripes are observed, and the intensity does not seem to vanish at the walls of the waveguide. This is an experimental artifact due to a lag of the perturbation body behind the guiding magnet during the measurements. Furthermore, the intensity distributions do not show the expected symmetry. This again is an experimental artifact resulting from the choice of the rectangular grid of positions of the perturbation body, one side being parallel to, respectively, the lower lead of the waveguides in the figure.}
\label{fig:WFexamples}
\end{figure}

To obtain a better understanding of the resonances their electric field intensity distributions were measured using the perturbation body method \cite{Maier1952a, Maier1952}. A cylindrical perturbation body with diameter $2$~mm and height $2.6$~mm made of a rubberlike plastic combined with magnetic barium ferrite powder
was introduced into the cavity and moved by an external guiding magnet to different positions $(x, y)$ on a grid with $2.5$~mm spatial resolution. For each position of the perturbation body, the shift $\Delta f$ of the resonance frequencies was measured. It is proportional to the electric field intensity~\cite{Bogomolny2006}, $I\propto |E_z(x,y)|^2$, at the position of the perturbation body,
\be \Delta f\propto |E_z(x, y)|^2 \, . \ee

The measured intensity distributions at the frequencies of the resonances with labels (a)--(d) in \reffig{fig:freqSpec} are plotted in \reffig{fig:WFexamples}. In panels (a)--(c) they exhibit yellow (bright) stripes, the color of which does not change to blue (dark) towards the walls of the waveguide, that is, the intensity does not seem to vanish there. This is an experimental artifact resulting from the friction between the perturbation body and the surface of the brass waveguides, which can lead to a lag of it behind the guiding magnet yielding an inaccuracy in its position. 

The intensity distributions shown in panels (a), (b) and (c) of Fig.~\ref{fig:WFexamples} are localized in the corner region of the waveguide and decay in the leads of the waveguide.
The resonances (a) and (b) evidently correspond to bound states as expected, since their frequencies are well below the cut-off frequency. The frequency of resonance (c) lies within the gray bar indicating the range of the expected value of $\fco$ (see inset of \reffig{fig:freqSpec}). Its intensity distribution, however, is localized in the corner region, which demonstrates that it also is a bound state. Still, it extends significantly into the leads. This is further discussed in \refsec{sec:expAdd}. Note that the length $l$ of the leads was chosen such that it is much larger than the decay lengths of the electric field intensities associated with the bound states. 

In contrast to the intensity distributions associated with the resonances (a)--(c), that corresponding to the resonance (d) extends over the whole waveguide and shows a periodic behavior in the leads [see \reffig{fig:WFexamples} (d)]. This is expected since its resonance frequency is larger than the cut-off frequency $\fco$ for propagating modes inside the leads. The resonance itself is very broad with $Q \approx 200$ since microwave power leaks out at the open ends of the waveguide, i.e., it corresponds to an unbound state. Figures~\ref{fig:WFexamples}(a)--(d) demonstrate that bound and unbound states can be clearly distinguished by means of their intensity distributions. 

\begin{figure}[tb]
\begin{center}
\includegraphics[width = 8.4 cm]{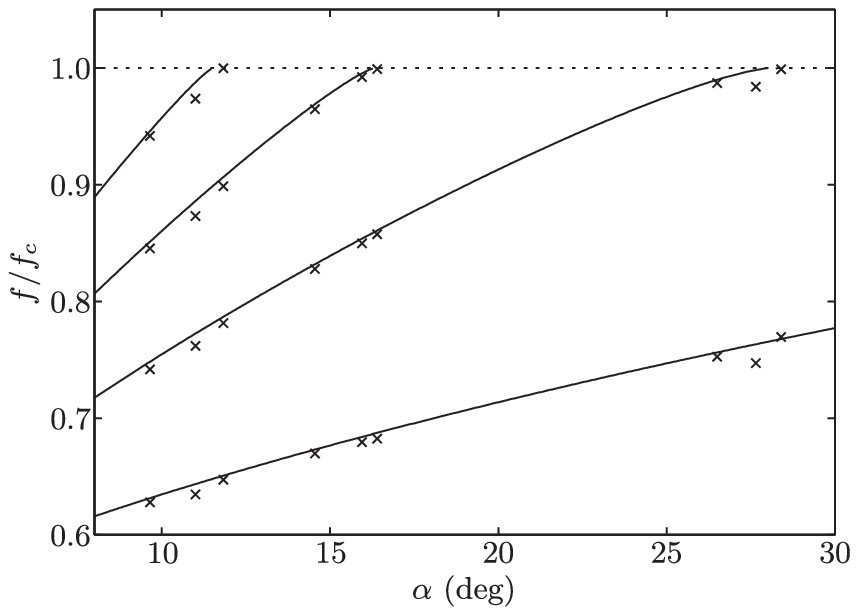}
\end{center}
\caption{Measured (crosses) eigenfrequencies of the bound states in units of the cut-off frequency in their dependence on the bending angle $\alpha$. The solid lines correspond to the eigenfrequencies computed in \refsec{sec:num} with a conventional numerical method (same data as in \reffig{sfig:Num1b}, but shown only in the range $8^\circ\leq \alpha\leq 30^\circ$). When an eigenfrequency crosses the dashed line, it reaches the value of the cut-off frequency. The occasional, slight deviations between the experimental and the numerical results are attributed to imperfections in the construction of the waveguides resulting from the soldering process. Also small differences between the critical angles computed in \refsec{sec:num} and the experimental ones are observed. }
\label{fig:expCalcFreqs}
\end{figure}

Figure~\ref{fig:expCalcFreqs} shows the measured eigenfrequencies in units of the cut-off frequency $\fco$ of the bound states (crosses) together with the values calculated in \refsec{sec:num} with a conventional method (solid lines, cf.~\reffig{Num1}). The agreement of both is good except for small deviations which are attributed to imperfections in the construction of the waveguides resulting from the soldering process. 

It should be noted that all the measured frequencies are below $\fco$, even though the bending angles $\alpha = 28.40^\circ$, $16.40^\circ$ and $11.83^\circ$ are slightly above the corresponding critical angles marked by the arrows in Fig.~\ref{sfig:Num1b}. However, in all three cases the frequency of the highest bound state is within the error range $\Delta \fco = 12$~MHz of the cut-off frequency indicated by the gray bar in~\reffig{fig:freqSpec}, and the bending angle of the waveguides is only known with a precision of $\Delta \alpha = 0.1^\circ$. In conclusion, within these uncertainties the measurements confirm the values of the critical angles deduced from the numerical calculations of \refsec{sec:num}.

\section{\label{sec:theo} Effective potential for sharply bent waveguides}
In this section we present an analytical approach to determine the bound states in the waveguides and the values of $\acrits$. It involves the solution of the Schr{\"o}dinger equation for a Hamiltonian with an effective binding potential, induced by a conformal map, based on an extension of the WKB method to systems comprising sharp corners~\cite{Bestle1995}. A first work on this subject was published in Ref.~\cite{Sadurni2010}, which unfortunately contains misprints. For this reason we revisit the approach and derive an improved approximation for the effective binding potential. 

\subsection{The conformal map}

The essential idea of the theoretical approach in Ref.~\cite{Sadurni2010} was to transform the Schr{\"o}dinger equation~(\ref{eq1}) with the coordinates $x,y\in\Omega$ confined to the interior of a sharply bent waveguide to one with coordinates $0\leq u\leq 1$, $-\infty <v<\infty$ restricted to an infinite strip. One procedure to obtain such a transformation is to introduce complex variables $z=x+iy$ and $\xi=u+iv$ and a transformation $z=F(\xi)$, which maps the infinite straight strip in the complex plane defined by the range of values of $u$ and $v$ onto the interior of the bent waveguide in the $(x,y)$ coordinate system. 

There are many possibilities to define $F(\xi)$. However, we impose the requirements that it should be smooth everywhere except at the corners of the bent waveguide and that the Laplacian $\Delta_{x,y}=\frac{\partial^2}{\partial x^2}+\frac{\partial^2}{\partial y^2}$ should be transformed into a Laplacian $\Delta_{u,v}=\frac{\partial^2}{\partial u^2}+\frac{\partial^2}{\partial v^2}$ without cross terms. 

For $\alpha >0^\circ$ a transformation with these properties was constructed in Ref.~\cite{Sadurni2010} using a combination of two Schwarz-Christoffel mappings~\cite{Schinzinger2003}, 
\be
\xi\rightarrow\chi=\sin^2\left(\frac{\pi}{2}\xi\right)\rightarrow z=F(\xi)
\label{map}
\ee
which images the infinite strip $\xi=u+iv$ onto
\be
\begin{array}{rcl}
z = F(\xi) & = & \frac{1}{\pi}\int_0^\chi{\rm d}t \, t^{\frac{\tilde\alpha}{2}-1}(1-t)^{-\frac{\tilde\alpha}{2}} \\
 & = & \frac{1}{\pi}B\left(\chi,\frac{\tilde\alpha}{2},1-\frac{\tilde\alpha}{2}\right),
\end{array}
\label{ConfMap}
\ee
where $\tilde\alpha = \frac{\alpha}{180^\circ}$ and $B(x,p,q)$ is the incomplete Beta function~\cite{Gradshteyn2007}. 

The image $z(u,v)=F(\xi)$ of the point $(u,v)=(0,0)$ is the outer corner of the bent waveguide at $(x, y)=(0, 0)$, that of the point $(u, v)=(1, 0)$ is its inner corner at $(x,y)=(\csc [\alpha/2],0)$, while the line $(0, v)$ is mapped onto its outer wall and the line $(1, v)$ onto its inner wall. 

\begin{figure}[bt]
\begin{center}
\includegraphics[width = 8.4 cm]{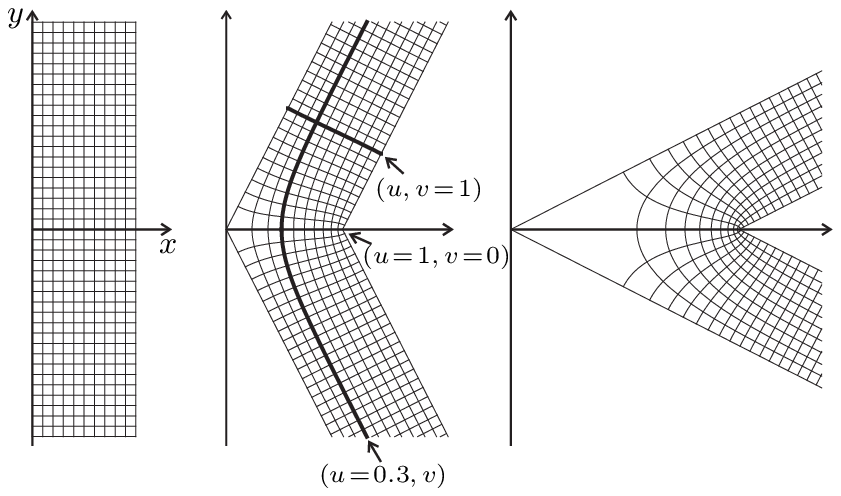}
\end{center}
\caption{\label{fig:confMap} Grid of the conformal coordinates $(x(u,v),y(u,v))$ obtained from the map $z = F(\xi)$ given in \refeq{ConfMap} for $\alpha = 180^\circ$, $126^\circ$ and $54^\circ$. For a fixed $v = \mathrm{const.}$ the coordinate lines are perpendicular on the boundaries $(x(u = 1, v),y(u = 1, v))$ and $(x(u = 0, v),y(u = 0, v))$, while for fixed $u = \mathrm{const.}$ the coordinate lines are parallel to the latter for large values of $v$. An abrupt change of the contour density occurs near the corners which is related to the Jacobian of the transformation $z = F(\xi)$ as expressed by \refeq{Jacobian_exp}. This feature stands outmost clearly in the example shown on the right with an accumulation of lines at the inner corner and a blank space to the right of the outer corner.}
\end{figure}

Figure~\ref{fig:confMap} shows three bent waveguides with angles $\alpha = 180^\circ$, $126^\circ$ and $54^\circ$, together with the coordinate grid $(x,y)=(x(u,v),y(u,v))$. For a fixed $v = \mathrm{const.}$ the coordinate lines approach the boundaries along $(x(0,v),y(0,v))$ and $(x(1,v),y(1,v))$  perpendicularly, whereas for a fixed $u = \mathrm{const.}$ they are parallel to the boundaries for large values of $v$. In Fig.~\ref{fig:confMap} examples are shown for both types of coordinate lines. 

\subsection{The Jacobian}
The change of coordinates implies the transformation
\be
\Delta_{x,y}=\frac{\partial (u,v)}{\partial (x,y)}\Delta_{u,v},
\ee
where the inverse of the Jacobian 
\be
\mathcal{J}(u,v;\tilde\alpha)=\frac{\partial (x,y)}{\partial (u,v)}\equiv\left\vert\frac{{\rm d}z}{{\rm d}\xi}\right\vert^2
\label{Jacobian}
\ee
appears as a prefactor and the Schr{\"o}dinger equation~(\ref{eq1}) takes the form 
\be 
\label{eq2} -\frac{\partial(u, v)}{\partial(x, y)}\Delta_{u, v}\psi(u, v)= E\psi(u, v)\ee
with
$\psi(u,v)\equiv\phi[x(u, v),y(u, v)]$ and
\be\psi(0, v) = \psi(1, v) = 0. \ee

To compute the Jacobian we perform the variable transformation $t=\sin^2\left(\frac{\pi}{2}\xi\theta\right)$ in the integral of \refeq{ConfMap} and obtain the expression
\be
z=\xi\int_0^1{\rm d}\theta\left[\cot\left(\frac{\pi}{2}\xi\theta\right)\right]^{1-\tilde\alpha}
\ee
which with \refeq{Jacobian} yields
\bea
\mathcal{J}(u,v;\tilde\alpha)=&\left[\left\vert\cot\left(\frac{\pi}{2}\xi\right)\right\vert^2\right]^{1-\tilde\alpha}\\
=&\left(\frac{\cosh(\pi v) + \cos(\pi u)}{\cosh(\pi v) - \cos(\pi u)}\right)^{1-\tilde\alpha}.
\label{Jacobian_exp}
\eea
For $\tilde\alpha=1$, i.e., $\alpha = 180^\circ$ we find $\mathcal{J}(u,v;\tilde\alpha=1)=1$ and the Schr{\"o}dinger equation reduces to that of a straight waveguide, the solutions of which are known. 

The abrupt change of the contour density observed in \reffig{fig:confMap} for $\alpha < 180^\circ$ close to the corners is related to the behavior of the Jacobian in their vicinity. In fact, near the outer corner, where $\vert\xi\vert\simeq 0$, the Jacobian can be approximated as 
\be\mathcal{J}(u,v;\tilde\alpha)\simeq\left(\frac{2}{\pi\vert\xi\vert}\right)^{2-2\tilde\alpha},\ee 
that is, it has a singularity of the type $x^{-\gamma}$ with the order $0 \leq \gamma = 2-2\tilde\alpha \leq 2$. On approaching the inner corner, i.e., for $\vert\xi-1\vert\rightarrow 0$, 
\be\mathcal{J}(u,v;\tilde\alpha)\simeq\left(\frac{\pi\vert\xi -1\vert}{2}\right)^{2-2\tilde\alpha}\ee 
becomes vanishingly small. For large values of $v\gg 1$ the Jacobian can be approximated as 
\be\mathcal{J}(u,v;\tilde\alpha)\simeq 1+2\left(1-\tilde\alpha\right)e^{-\pi v}\cos(\pi u)\ee 
and converges to unity. This behavior is expected since there the transformation from $\xi$ to $z=F(\xi)$ simply corresponds to a rotation of the coordinate grids by an angle $\frac{\pi}{2}\left(1-\tilde\alpha\right)$ with respect to each other around the inner corner.

\subsection{Emergence of the potential}

Multiplication of the Schr{\"o}dinger equation (\ref{eq2}) with the Jacobian $\mathcal{J}(u,v;\tilde\alpha)$ yields
\be
\left(-\Delta_{u,v}+\left[1-\mathcal{J}(u,v;\tilde\alpha)\right]E\right)\psi(u,v)=E\psi(u,v),
\label{effPot}
\ee
that is, the mapping \refeq{map} entails the emergence of an effective binding potential depending on the Jacobian $\mathcal{J}(u,v;\tilde\alpha)$, i.e., the area density. 
To obtain an approximate analytical solution of this boundary value problem it is turned into an effective potential problem in the $v$ coordinate by choosing an appropriate orthonormal basis for the $u$ coordinate. The $u$-coordinate lines approach the $v$-coordinate lines at the boundaries of the waveguide perpendicularly (see \reffig{fig:confMap}). Thus a suitable choice that obeys the boundary conditions imposed on $\psi(u,v)$ [see \refeq{eq2}] is given by
\be
\label{eq4} \psi(u,v)=\sqrt{2}\sum_{n=1}^{\infty}f_{n}(v)\sin\left(n\pi u\right) 
\ee
with
\be
f_{n}(v)=\sqrt{2}\int_{0}^{1}{\rm d}u \, \psi(u,v)\sin\left(n\pi u\right) \, .
\ee

Furthermore, since the only singularity of the Ja\-co\-bian, which is located at $(u,v) = (0,0)$, has an order less than $2$, products of the form $\sin(m\pi u)\sin(n\pi u)\mathcal{J}(u,v;\tilde\alpha)$ are regular functions at all points inside the waveguide. Therefore, we may expand the function $\psi(u, v)$ in \refeq{effPot} in terms of the orthonormal functions $\sqrt{2}\sin(n\pi u),\, n=1,2,\dots$ and obtain the system of coupled equations
\be\label{eq6} 
\sum_{m=1}^{\infty}\left[-\delta_{nm}\partial^2_{v}+E\tilde V_{nm}(v)\right]f_{m}(v)=E_nf_n(v)
\ee
with 
\be\label{eq5}
E_n=E-\left(n\pi\right)^2,\, \tilde V_{nm}(v)=\delta_{nm}-V_{nm}(v), 
\ee
where
\be\label{eq5a} 
V_{nm}(v)= 2\int_{0}^{1}{\rm d}u\sin(n\pi u)\sin(m\pi u)\mathcal{J}(u,v;\tilde\alpha)
\ee 
are the coupling matrix elements. In this equation the energy $E$ appears parametrically as the strength of the coupling matrix elements  $V_{nm}(v)$ which define an infinite-dimensional matrix. The numerical evaluation of the integral \refeq{eq5a} yields three characteristic properties of $V_{nm}(v)$ which may also be deduced from those of the conformal map: $(i)$ the matrix elements $V_{nm}(v)$ decrease as one moves away from the diagonal, $|V_{n,n+m}(v)|\approx |V_{nn}(v)|e^{-mv},\, m=1,2,\cdots$ for all $v$, $(ii)$ the diagonal elements saturate with increasing $n$, i.e., $V_{nn}(v) \leq\int_{0}^{1}{\rm d}u\mathcal{J}(u, v;\tilde\alpha)$, and (iii) for $v \rightarrow \infty$ the matrix $V_{nm}$ approaches the identity matrix, $V_{nm}(v) \rightarrow \delta_{nm}$, i.e., $\tilde V_{nm}(v) \rightarrow 0$. 
\subsection{Effective potential}
Numerical tests revealed that for small values of $v\lesssim 0.25$ we may write to a good approximation
\be
\sum_{m=1}^{n-1}\left(V_{nm}(v)+V_{mn}(v)\right)+V_{nn}(v)\simeq nV_{nn}(v),\label{approx}
\ee
where the exact range of validity depends on $n$ and on $\alpha$, whereas for $v\gtrsim 1.5$
\be
V_{nm}(v)\simeq V_{nn}(v)\delta_{nm}.
\ee
Accordingly, for small and large values of $v$ we may replace Eq.~(\ref{eq6}) by 
\be\label{eq6a}
\left(-\partial^2_{v}+E\tilde V_n(v)\right)f_{n}(v)=E_nf_n(v).
\ee
with
\begin{eqnarray}
&\tilde V_n(v)=1-nV_n(v)\, {\rm for}\, v\lesssim 0.25,\label{eq6b}\\
&\tilde V_n(v)=1-V_n(v)\, {\rm for}\, v\gtrsim 1.5,\label{eq6c}
\end{eqnarray} 
where we use the notation $V_n(v)=V_{nn}(v)$.
\begin{figure}[bt]
\begin{center}
\subfigure[]{
        \includegraphics[width = 8.4 cm]{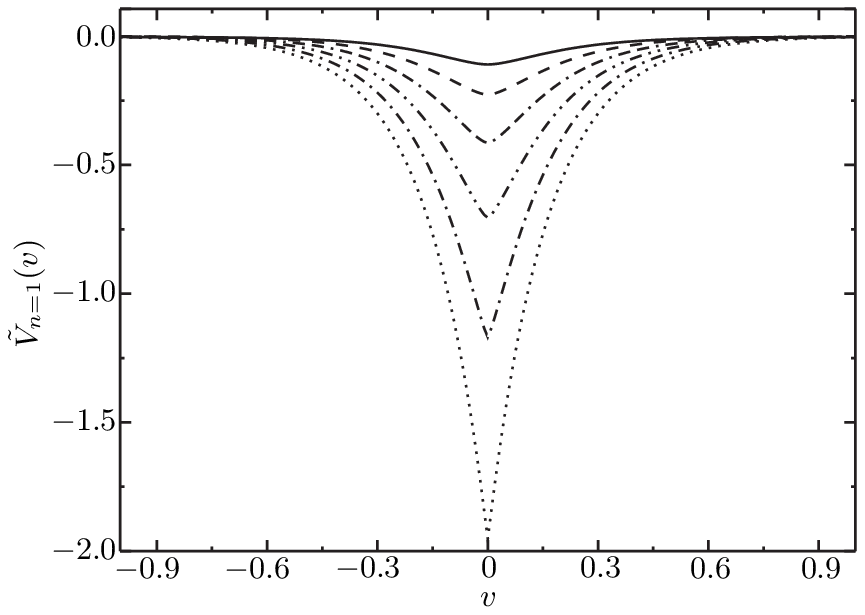}
	\label{Wkb1}
}
\subfigure[]{
       \includegraphics[width = 8.4 cm]{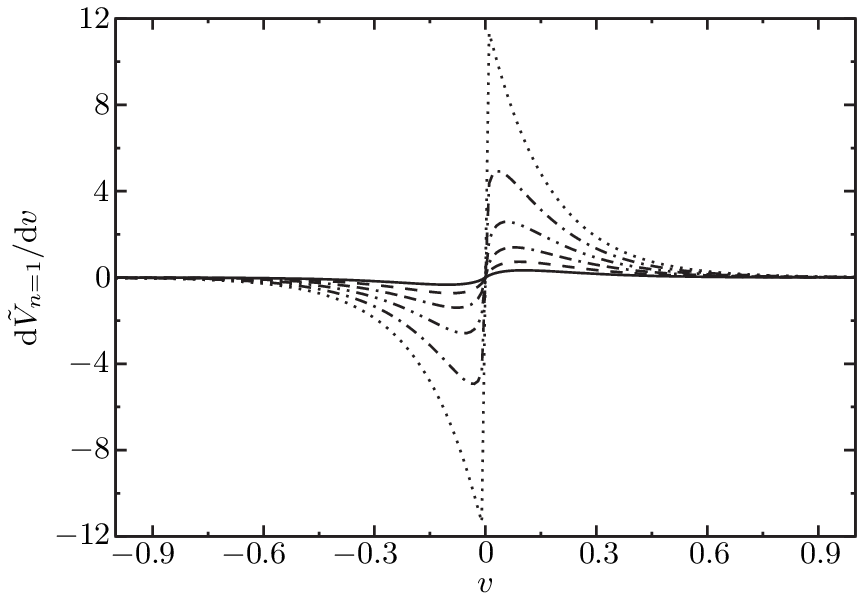}
	\label{Wkb2}
}
\end{center}
\caption{\label{figpotenzial} \subref{Wkb1} Effective potential $\tilde V_{n=1}(v)=1-V_{n=1}(v)$ and \subref{Wkb2} the derivative $\frac{{\rm d}\tilde V_{n=1}(v)}{{\rm d}v}=-\frac{{\rm d}V_{n=1}(v)}{{\rm d}v}$ for $\alpha =1^\circ$ (dotted line), $\alpha =25^\circ$ (dashed-dotted line), $\alpha =49^\circ$ (dashed-dotted-dotted line), $\alpha =73^\circ$ (dashed-dashed-dotted line), $\alpha =97^\circ$ (dashed line) and $\alpha =121^\circ$ (full line).}
\end{figure}

In terms of a Hamiltonian $H(v)$ the Schr\"odinger equation would read
\be
\left[H(v)- E_n\right]f_{n}(v)=0,
\ee
with the eigenenergies $E_n$ defined in Eq.~(\ref{eq5}). They are related to the eigenvalues $E$ of Eq.~(\ref{eq2}) which appear in Eq.~(\ref{eq6a}) as the prefactor of the potential $\tilde V_n(v)$.

The coupling matrix elements defined by \refeq{eq5a} depend on the bending angle $\alpha$. Figures~\ref{Wkb1} and~\ref{Wkb2} show $\tilde V_{n=1}(v)=1-V_{n=1}(v)$ and its derivative $-\frac{{\rm d}V_{n=1}(v)}{{\rm d}v}$ for six different values of $\alpha$. The potential $\tilde V_n(v)$ is a symmetric function of $v$, has a minimum at $v = 0$ and approaches $0$ for $|v| \gtrsim 1.5$. Furthermore it has the shape of a kink for $\alpha\lesssim 30^\circ$. Similarly for decreasing $\alpha$ its derivative, though vanishing at $v=0$, undergoes an increasingly rapid change from its minimum $V^\prime_{\rm min}=-\frac{{\rm d}V_n(-\epsilon
 )}{{\rm d}v}$ to its maximum $V^\prime_{\rm max}=-\frac{{\rm d}V_n(\epsilon)}{{\rm d}v}$ situated at $v=\pm\epsilon$ with $\epsilon\ll 1$, which is comparable to a jump $\kappa=V^\prime_{\rm max}-V^\prime_{\rm min}$ for small bending angles $\alpha$. 

\subsection{Semiclassical approximation}
In Ref.~\cite{Bestle1995} an extension of the WKB approximation to systems described by a potential which displays a corner and has a discontinuous derivative is proposed. We now apply this approach to the present situation and determine approximately the bound states. 

\subsubsection{Quantization condition}
Due to the minimum of the effective potential, bound states may exist for $E_n\leq E\tilde V_n(v)\leq 0$. The value of $v$, where equality $E_n= E\tilde V_n(v)$ holds, defines the turning point $v_t=\vert v\vert$. For $n=1$ the inequality is not satisfiable for $E\geq\pi^2$. This value defines the cut-off energy $E_c$ above which no bound states exist. For $E=0.9999\pi^2$ we obtain $v_t\simeq 1.5$ for $n=1$. For $n\geq 2$ the value of the turning point is smaller than $0.25$. Furthermore, the extended WKB approximation involves the potential around the corner at $v=0$. Accordingly, we may replace $\tilde V_n(v)$ by $1-nV_n(v)$ in Eq.~(\ref{eq6a}) using Eq.~(\ref{eq6b}). 

The corner-corrected WKB approximation yields for a symmetric potential the quantization condition~\cite{Bestle1995}
\be\label{wkb_eq} 
\frac{\Delta}{4}=\tan\left[\pi\left(\lambda +\frac{1}{4}\right)-S\right],\ee
where we defined the action
\be S=\int_0^{v_t}{\rm d}v\sqrt{(n\pi)^2-EnV_n(v)},\ee
and introduced the notations
\be \label{eq:WKB_Delta} \Delta = \left[ (n\pi)^2-EnV_n(0) \right]^{-3/2} \frac{\kappa}{2}nE \, , \ee
and $\lambda=0, 1, 2, \dots$. Since the solutions of \refeq{wkb_eq} do not depend on the value of $\lambda$ we may set $\lambda = 0$. 

Although the explicit forms of the potential and its derivative, which is needed for the computation of $\kappa$, are not available, the associated integrals \refeq{eq5a} and thus \refeq{wkb_eq} can be solved numerically. For simplicity we define the rescaled energies $e=\frac{EV_n(0)}{\pi^2}$ and \refeq{eq:WKB_Delta} takes the form 
\be 
\left(n-e\right)^{3/2}=\frac{\kappa}{2V_n(0)}\frac{e}{4\pi\sqrt{n}}\frac{\cos(S)+\sin(S)}{\cos(S)-\sin(S)}.\label{wkb_neu}
\ee 

For $n=1,2,\dots, \mathcal{N}_b(\alpha)$, with $\mathcal{N}_b(\alpha)$ denoting the total number of bound states at $\alpha$, this equation yields the rescaled eigenvalues $e_n(\alpha)$ of the bound states of a bent waveguide. Here, the values of $e_n(\alpha)$ increase with their index, i.e., $e_n(\alpha)\leq e_{n+1}(\alpha)$. They are bounded from above by the condition $e\leq \mathcal{N}_b(\alpha)$. Beyond that value all states are unbounded. Thus, the cut-off value of the rescaled energies is given by $e_c(\alpha)=\mathcal{N}_b(\alpha)$. 
\subsubsection{Number of bound states}
We determined the number of bound states $\mathcal{N}_b(\alpha)$ with the help of~\cite{Calogero1967,Landau1977}.
\be
\left[\mathcal{N}_b(\alpha)-1\right]=\frac{\int{\rm d}u{\rm d}p_u}{\pi}\frac{\int{\rm d}v{\rm d}p_v}{2\pi}=\frac{1}{\pi}\int_0^1{\rm d}u\int_{-\infty}^{\infty}{\rm d}v\frac{p_{\rm max}^2}{2}.
\ee
The value of $p_{\rm max}$ is obtained from the classical analog of Eq.~(\ref{effPot}), which can be written as $E_{\rm class}=\frac{p^2}{2}+\left(1-\mathcal{J}(u,v;\tilde\alpha)\right)\frac{E}{\pi^2}$, and satisfies the inequality $E_{\rm class}\leq 0$ for the bound states. Thus, $\frac{p_{\rm max}^2}{2}=\left(\mathcal{J}(u,v;\tilde\alpha)-1\right)\frac{E_c}{\pi^2}$, yielding
\be
\mathcal{N}_b(\alpha)=\frac{1}{\pi}\int_0^1{\rm d}u\int_{-\infty}^{\infty}{\rm d}v\left(\mathcal{J}(u,v;\tilde\alpha)-1\right)+1=\frac{\tilde\mathcal{A}(\alpha)}{\pi}+1.
\ee
Note, that for finite lead lengths, i.e., for $v$ restricted to some value $v\leq v_{\rm max}$, $\int_0^1{\rm d}u\int_{-v_{\rm max}}^{v_{\rm max}}{\rm d}v\left(\mathcal{J}(u,v;\tilde\alpha)-1\right)=\mathcal{A}(\alpha)-2v_{\rm max}$, where integration over the Jacobian, $\mathcal{A}(\alpha)$, yields the area of the finite bent-waveguide. The integral already approaches the value $\tilde A(\alpha )$ for $v_{\rm max}\simeq 1.5$. In fact, $\tilde A(\alpha)$ can be interpreted as the effective area covered by the potential, as it corresponds to the area of an unbounded bent waveguide minus the area of the leads where the potential becomes vanishingly small. In the interval considered, $0.7^\circ\leq\alpha\leq 30^\circ$, it is well described by $\tilde\mathcal{A}(\alpha)=\csc(\alpha /2)-a_0$, with $a_0\simeq 0.91$, that is, to a good approximaton the integer part of $\mathcal{N}_b(\alpha)$ coincides with $\mathcal{N}(\alpha)$ given in Eq.~(\ref{fitnumber}). It equals $0$ for $\alpha =180^\circ$.

\subsubsection{Approximate critical angles and rescaled energies}
For $\alpha <180^\circ$ the rescaled energies, i.e., the ratio of the solutions $e_n(\alpha)$ of \refeq{wkb_neu} and the cut-off energy $e_c(\alpha)$ is given by 
\be\tilde e_n(\alpha) = \frac{1}{\mathcal{N}(\alpha)} e_n(\alpha).\label{cutoff}\ee 
\begin{figure}[bt]
\begin{center}
\subfigure[]{
        \includegraphics[width = 8.4 cm]{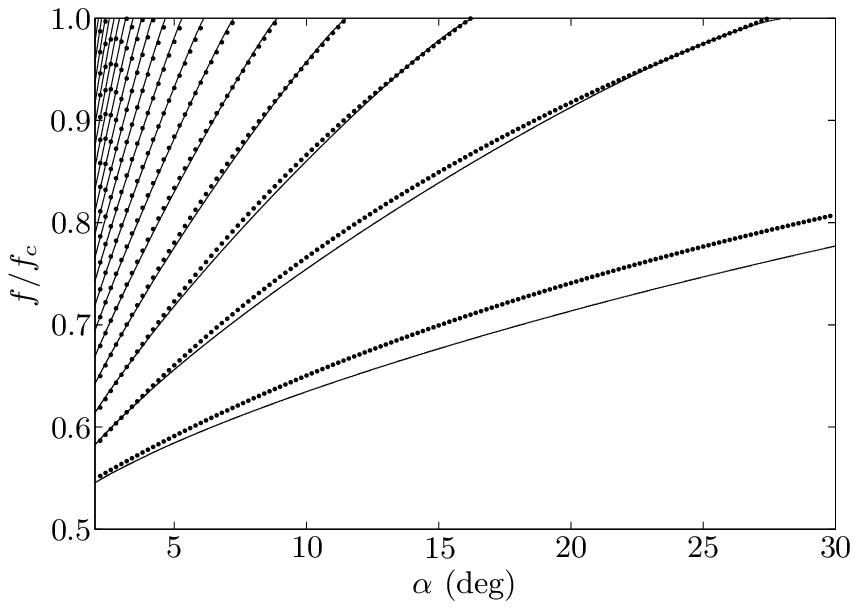}
        \label{sfig:NumWKBa}
}
\subfigure[]{
        \includegraphics[width = 8.4 cm]{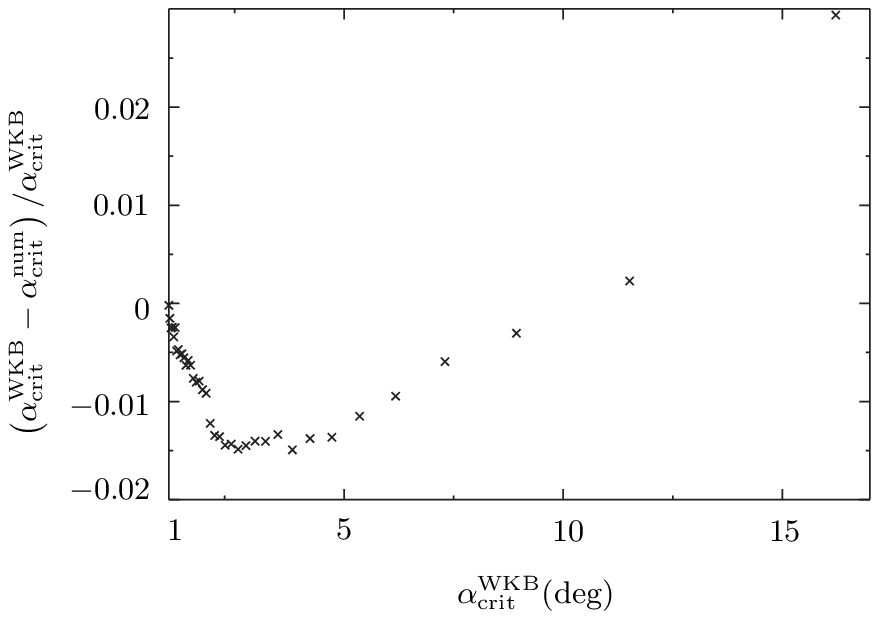}
        \label{sfig:NumWKBb}
}
\end{center}
\caption{\label{fig:NumCalc} \subref{sfig:NumWKBa} Comparison between the rescaled eigenfrequencies $f/f_c$ of the bound states computed with the conventional numerical method outlined in~\refsec{sec:num} (full lines) with those obtained from the corner-corrected and ground-state shifted WKB approximation (dots). \subref{sfig:NumWKBb} Relative deviations of the critical angles evaluated numerically and with the WKB approximation versus $\acrits^{\rm WKB}$. Both panels demonstrate that the agreement between the numerical and the WKB results is already good for the third bound state. }
\end{figure}

Generally the dependence of the solution of \refeq{wkb_neu} on $S$ is negligibly small for $\alpha\gtrsim 1^\circ$, so the last term in this equation can be set to unity. Then the solutions of \refeq{wkb_neu} are given as the zeros of a polynomial of third order. These are computed with the help of Cardano's method~\cite{Korn1968}. 

To determine the critical angles, where a bound state turns into an unbound one, we furthermore have to take into account that Eq.~(\ref{wkb_neu}) does not yield the correct ground state energy~\cite{Bestle1995}. Indeed, for $\alpha\rightarrow 0^\circ$ the rescaled eigenenergy $\tilde e_1(\alpha) = e_1(\alpha) / e_c(\alpha)$ does not converge to the expected value $e / e_c = f^2 / f_c^2 = 0.25$ (see \reffig{Num1}). However, when introducing a constant shift $e_0 \approx 0.25$ of the rescaled energies $\tilde e_n(\alpha)$, which ensures the correct convergence for $\alpha \rightarrow 0^\circ$, a good agreement between the solutions $\acrits$ of the equation $e_0 + \tilde e_n(\acrits) = 1$ and the numerical results of \refsec{sec:num} and thus also the experimental ones of \refsec{sec:exp} is achieved. This property is also true for the bound states as demonstrated in Fig.~\ref{sfig:NumWKBa}. Already for the second bound state the agreement between the numerical values of $f / f_c$ of \refsec{sec:num} and those obtained with the corner-corrected and ground-state shifted WKB approximation is good. 

The relative deviations of the critical angles obtained with the two methods are shown in \reffig{sfig:NumWKBb}. They are largest for large critical angles, however less than $2\%$, and decrease rapidly to  a relative deviation of less than $1\%$ for $\acrits\lesssim 1^\circ$. Thus the WKB approximation proposed in Ref.~\cite{Bestle1995} for systems with a potential displaying a kink provides a good prediction of the critical angles. However, it is necessary to include the known result for the behavior of the rescaled energies, respectively, frequencies in the limiting case $\alpha \rightarrow 0^\circ$.

\subsection{Summary}

In summary, the conventional numerical methods, also the one presented in \refsec{sec:num}, used to compute the eigenfrequencies and wave functions of the bound states involve the solution of the Schr\"odinger equation of a free particle with Dirichlet boundary conditions, \refeq{eq1}. In contrast, in this section a Schr\"odinger equation for a Hamiltonian including an effective potential was derived, the spectrum of which contains the eigenfrequencies of the bound states in a bent waveguide. These are determined by solving the associated Schr\"odinger equation (\ref{eq6a}) with a semiclassical approximation. A comparison with the numerical results of \refsec{sec:num} and the experimental ones of \refsec{sec:exp} corroborates the validity of the theoretical approach in terms of an effective binding potential presented in this section. The property of the effective potential that it is nonvanishing for $v\lesssim v_{\rm max}$ implies that its influence reaches into the leads of the waveguide, in accordance with the penetration of the wave functions into the leads as observed in the calculated intensity distributions in \reffig{Num3} and the experimental ones in \reffig{fig:WFexamples}. This feature already indicates that the existence of a bound state should depend on the length of the leads of the waveguide. The results of an experimental investigation of the effect of the finite length are presented in the next section. 

\section{\label{sec:expAdd}Finite leads and resonant tunneling: experimental investigation}

The numerically calculated intensity distributions and the experimental ones in Figs.~\ref{Num3} and \ref{fig:WFexamples}, respectively, clearly demonstrate that there is a leakage of the bound state wave functions from the interior region into the leads of the waveguide. This section focuses on the experimental investigation of the decay lengths of the bound state wave functions into the leads. 
\subsection{Decay into the leads}
While most theoretical works on bound states in bent waveguides assume leads of infinite lengths, this oversimplification of course is not feasible in the experimental realizations. Due to their leakage into the leads, modes localized in the corner region of a bent quantum wire or waveguide of finite length may couple to a continuum of states, e.g., to a 2d electron reservoir in the case of quantum wires \cite{Berggren1991, Wang1995, Wu1991, Wu1993} or to the electromagnetic waves in free space in the experiments presented in this paper. 

Similarly, waves sent into such a device may be trapped in the interior region before they are scattered back to the exterior. This phenomenon manifests itself in transmission spectra measured between both lead ends as resonances and is attributed to resonant tunneling via the bound states~\cite{Schult1989, Wu1991, Berggren1991, Wu1993, Wang1995, Carini1997b} which can significantly alter the transmittance characteristics of the devices. Strictly speaking, in a waveguide coupled to a continuum the associated states are no longer bound but acquire a finite lifetime since they can decay via tunneling through it. In this section we show some qualitative results on the effect of a finite lead length and investigate the transmission through the waveguides by attaching waveguide-to-coaxial adapters to the ends of the waveguides. 

\begin{figure}[tb]
\begin{center}
\includegraphics[width = 8.4 cm]{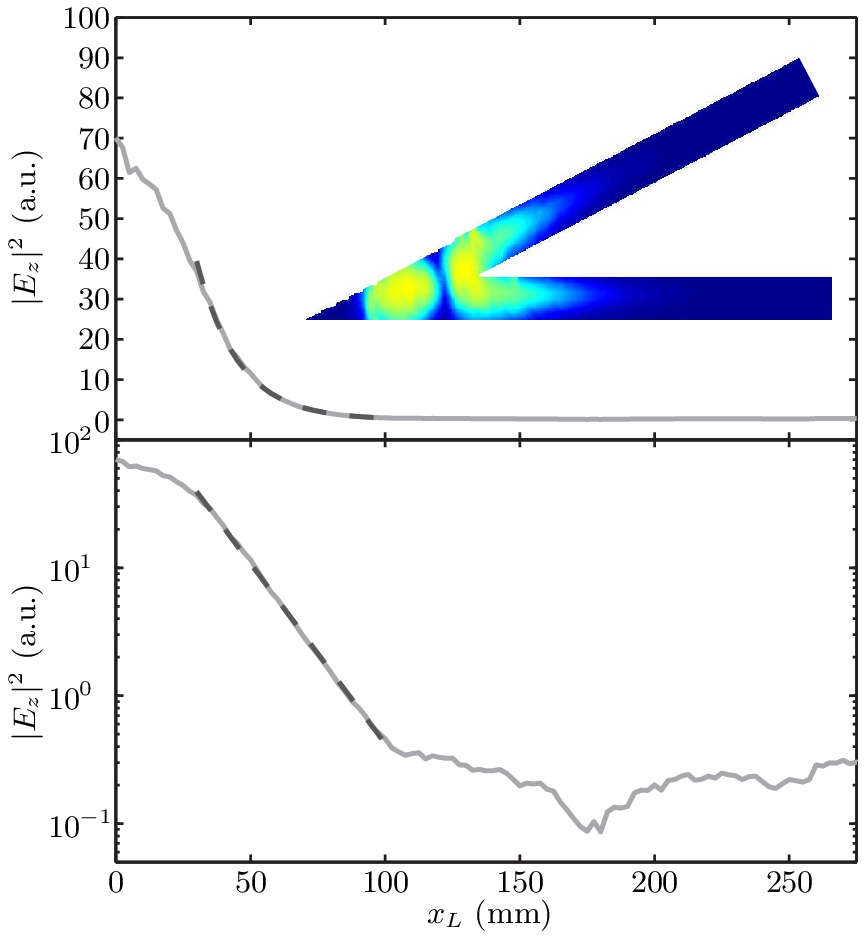}
\end{center}
\caption{(Color online) Decay of a bound state into the leads represented by the projection of the measured field distribution (gray line) onto the $x_L$-axis parallel to the waveguides [see \reffig{sfig:schemWG}]. Here we use the second bound state at $9.364$~GHz for $\alpha = 27.65^\circ$ with lead length $l_0$ and show this projection in linear scale (top panel) and logarithmic scale (bottom panel). In the regime from $20 {\rm mm}\leq x_L\leq 100 {\rm mm}$ the intensity distribution decays exponentially, indicated by the dashed line, with the decay length $\delta_\mathrm{expt} = (16.2 \pm 0.4)$~mm. The inset displays the complete intensity distribution but parts of the waveguides are not shown.}
\label{fig:decayLength}
\end{figure}

\begin{figure}[tb]
\begin{center}
\includegraphics[width = 8.4 cm]{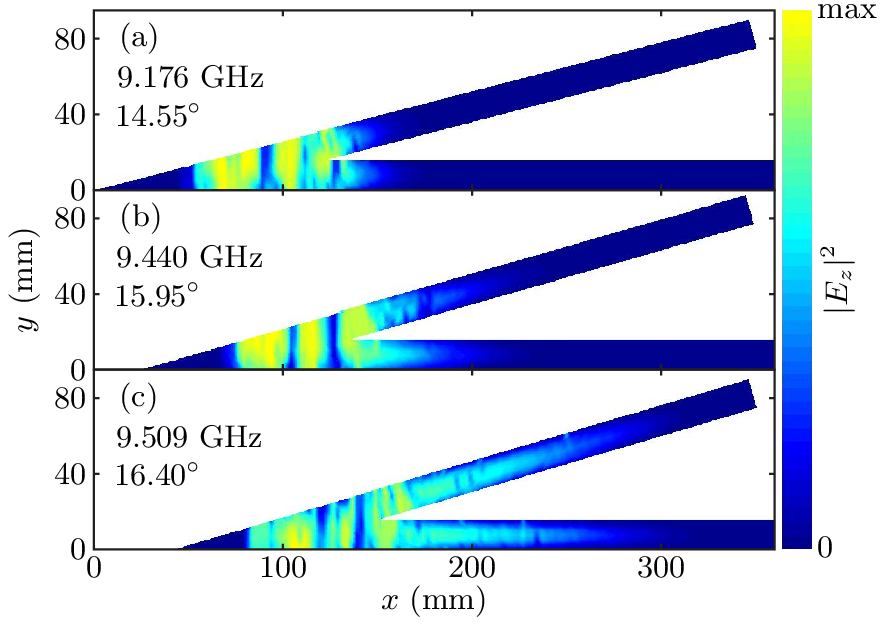}
\end{center}
\caption{(Color online) Influence of the short leads on the bound states as expressed by the leakage of the measured intensity distributions of the third bound state into the leads for angles (a) $\alpha = 14.55^\circ$, (b) $\alpha = 15.95^\circ$ and (c) $\alpha = 16.40^\circ$. The electric field intensity $I \propto |E_z|^2$ is plotted in false colors. The blue (darkest) color corresponds to the lowest, the yellow (brightest) to the highest intensity. In contrast to \reffig{fig:WFexamples}, the waveguides now have short leads with length $l_0$ summarized in~\reftab{tab:angles}. With increasing bending angle the intensity penetrates deeper into the leads.}
\label{fig:WFevolution}
\end{figure}

The measurements of the frequency spectra and the intensity distributions presented in \refsec{sec:exp} were repeated with waveguides without extensions, i.e., with lead lengths $l_0$ given in \reftab{tab:angles} and open ends (as in \refsec{sec:exp}). The decay lengths of the field intensity distributions were determined experimentally from the projection of the measured intensity distributions onto the coordinate $x_L$ parallel to the waveguides [see \reffig{sfig:schemWG}]. A so-obtained curve is depicted in \reffig{fig:decayLength}, where the measured intensity distribution of the second bound state of the waveguide with bending angle $27.65^\circ$ at $9.364$~GHz is shown. 

The projection of the intensity distribution onto $x_L$ is represented by a gray line in linear (top panel) and logarithmic scale (bottom panel). Three different regimes are visible in the logarithmic plot (bottom 
panel). The line is slightly curved for $x_L\lesssim 20$ mm, it is linear for $20{\rm mm}\lesssim x_L\lesssim 100$~mm, i.e., $|E_z|^2$ decays exponentially, \mbox{$|E_z|^2 \propto \exp{(-x_L/ \delta)}$}, in the leads of the waveguide and for $x_L \gtrsim 100$~mm the noise level is reached. The decay length was determined with a linear fit (dashed black line) in the middle part of the line as $\delta_\mathrm{expt} = (16.2 \pm 0.4)$ mm which agrees well with the value $\delta = 14.4$ mm computed from 
\be \label{eq:decayLength} \delta = \frac{c}{4 \pi \sqrt{\fco^2 - f^2}}. \ee
In this case, the eigenfrequency of the bound state is well below the cut-off frequency and the associated intensity distribution extends not very far into the waveguides. Generally, good agreement with Eq.~(\ref{eq:decayLength}) was found for states with eigenfrequencies $f\ll f_c$.

\subsection{Coupling to the continuum}
Figure~\ref{fig:WFevolution} shows the intensity distributions of the third bound state for waveguides with bending angles $14.55^\circ$, $15.95^\circ$ and $16.40^\circ$. As the bending angle $\alpha$ increases from panel (a)--(c), the resonance frequency approaches the cut-off frequency $\fco$ and the intensity distribution leaks more and more into the leads. This feature is in accordance with \refeq{eq:decayLength} since the decay length $\delta$ increases as $f$ approaches $\fco$. For $\alpha = 16.40^\circ$ it was determined as $\delta_\mathrm{expt} = 50.6$~mm from the measured intensity distribution, which is about one quarter of the lead length $l_0 = 211.4$~mm. Therefore, it can be expected that the finite lead length influences the properties of the state. Indeed, the measured resonance frequency, $9.509$~GHz, is \mbox{$8$~MHz} larger than that of the corresponding waveguide with extension~[see \reffig{fig:WFexamples}(c)] the length of which is $l=411.4$~mm~$\approx 2 l_0$. Such a shift by several MHz to larger frequencies when reducing the lead length was observed for most resonance states close to the cut-off frequency and was also predicted in Ref.~\cite{Delitsyn2012}. The quality factors $Q$ of the resonances, on the other hand, did not change significantly, thus indicating that the leakage due to tunneling is small. 

These results demonstrate that for the determination of the eigenfrequencies of the bound states the choice of sufficiently long waveguides is crucial. For the experiments presented in \refsec{sec:exp} the lead length $l$ of the waveguides with extension was chosen significantly larger than the decay lengths $\delta$ of the states with frequencies below $\fco$, in order to minimize the effect of their finite lengths. 

\begin{figure}[tb]
\begin{center}
\subfigure[]{
	\includegraphics[width = 6 cm]{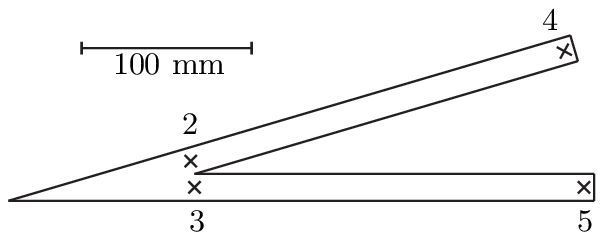}
	\label{sfig:setupTrans}
}
\subfigure[]{
       \includegraphics[width = 8.4 cm]{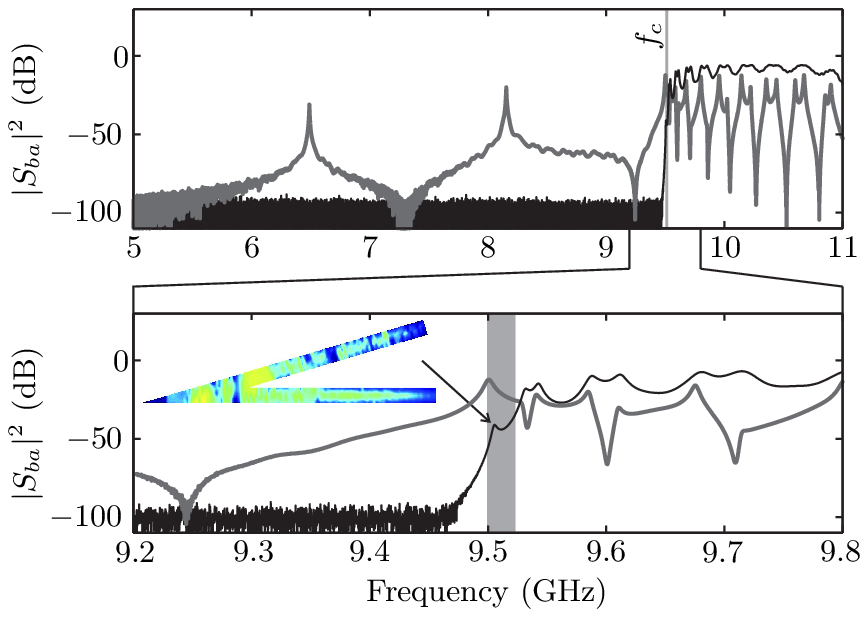}
	\label{sfig:spectrumTrans}
}
\end{center}
\caption{(Color online) Resonant tunneling in a bent waveguide observed in measured microwave transmission spectra. \subref{sfig:setupTrans} Sketch of the setup. The crosses marked $2$ and $3$ indicate the positions of the wire antennas also used in the experiments presented in \refsec{sec:exp} and the crosses marked $4$ and $5$ denote the positions of the waveguide-to-coaxial adapters. \subref{sfig:spectrumTrans} Frequency spectrum of the waveguide with $\alpha = 16.40^\circ$ and adapters attached. The spectrum was measured both with the adapters (black thin line) and with antennas $2$ and $3$ located near the tip of the waveguide (thick gray line). The light gray vertical bars and its width indicate the cut-off frequency $\fco$ and its error bar, respectively. The two resonances at $6.49$~GHz and $8.15$~GHz observed in $\vert S_{32}\vert^2$ are not seen in $\vert S_{54}\vert^2$ because the corresponding intensity distributions decay rapidly in the leads and thus cannot couple to the adapters. Only the third resonance below $f_c$ at $9.506$~GHz is observed in $\vert S_{54}\vert^2$, since its intensity distribution extends far into the waveguide (see inset in the bottom panel).}
\label{fig:transmissionSpectrum}
\end{figure}

Next, the transmission of microwave power from port $4$ at one lead end through the waveguides to port $5$ at the other one in Fig.~\ref{sfig:setupTrans} was measured in order to investigate the coupling of the bound states to the continuum. For this, waveguide-to-coaxial adapters (model 4609 by Narda Safety Test Solutions) were attached to the flanges at the ends of the waveguides (see \reffig{fig:expSetup}). The adapters close the system but provide a strong coupling to the attached coaxial cables. The geometry of this setup is shown in Fig.~\ref{sfig:setupTrans}. 
If the coupling to the exterior is non-negligible, then the bound states are observable as resonances below $\fco$ in the measured spectra like in Refs.~\cite{Wu1991,Wu1993,Dembo2004}.

The transmission spectrum obtained by coupling in and out microwave power via the two adapters is shown as black thin line in Fig.~\ref{sfig:spectrumTrans} for the waveguide with $\alpha =16.40^\circ$, that measured between the antennas $2$ and $3$ located in the leads close to the interior region as gray thick line. The latter exhibits three sharp resonances below $\fco$ at frequencies which were shown to correspond to bound states in \refsec{sec:exp} (see \reffig{fig:freqSpec}). The former, on the other hand, only shows one resonance below $\fco$, at $9.506$~GHz. The intensity distribution (inset in the bottom panel) measured at the corresponding frequency confirms that this resonance indeed can be assigned to the third bound state which apparently couples to the adapters. This can be attributed to its large decay length in the leads. Thus the coupling of the third bound state to the continuum indeed implies a resonance in the transmission spectrum due to resonant tunneling, whereas the first and second bound state do not. This is explainable with their considerably smaller decay lengths. 

Note that the resonance frequency of the third bound state differs slightly for the measurements with the antennas and the adapters, respectively. This is attributed to the fact that the coupling of the adapters to the waveguide is much stronger than that of the small wire antennas, as can be deduced from the average value of $|S_{ba}|^2$ above $\fco$, which is larger in the former case. 

\subsection{Summary}
Figure~\ref{sfig:spectrumTrans} shows that the bound states can indeed modify the transmission spectrum of the bent waveguides (see also Ref.~\cite{Carini1997b}) due to resonant tunneling as observed in quantum wires~\cite{Wu1991}. However, this was seen in just two cases for all nine waveguides, and only if no extensions for the leads were used. In both cases, the associated resonance frequency was less than $5$~MHz apart from the cut-off frequency. Apparently, for resonant tunneling to be observed, the decay length $\delta$ of the corresponding state must be about the same as the lead length. However, only states close to the cut-off frequency in waveguides with bending angles close to a critical angle satisfy this condition which emphasizes the importance of the knowledge of the critical angles. 

Furthermore our experiments show that especially for these cases the finite lead length and the details of the coupling can become important, which requires further theoretical investigations. First results regarding the influence of the lead length were published in Ref.~\cite{Delitsyn2012}.

\section{\label{sec:concl}Conclusions}

We have studied experimentally and numerically the occurrence of bound states in sharply bent waveguides below the threshold for propagating modes. In some of the experimental realizations the bending angle was chosen close to a critical angle. The measured eigenfrequencies of the bound states and the values of the critical angles were predicted to a high accuracy with two independent methods: The first method consists of the numerical solution of the Schr\"odinger equation for a free particle with Dirichlet boundary conditions at the walls of the waveguide based on a scattering approach. 

The second one encompasses the derivation of a Hamiltonian containing a binding potential by introducing a suitable conformal mapping. The associated Schr\"odinger equation was solved semiclassically using corner-corrected WKB approach~\cite{Bestle1995}. These results are of particular relevance for the transmittance of electromagnetic waveguides and quantum wires. The theoretical treatment is, to our knowledge, the first one to work out the problem by analytical methods, even though it relies on an approximation. The demonstration of its applicability to sharply bent waveguides renders possible the analytical treatment of two-dimensional problems with non-trivial boundaries through the use of a conformal mapping and an extended WKB approximation. 

In further experiments, the intensity distributions of the bound states were measured. They exhibit interesting features close to the cut-off frequency and for bending angles close to a critical one. Especially the influence of the finite lead lengths on their boundedness, that is, the size of their coupling to the exterior, was investigated. These studies revealed that predominantly bound states with resonance frequencies close to the cut-off frequency are affected if the length of the leads is of the same size as their decay length, the relevant mechanism being resonant tunneling.  

\begin{acknowledgments}
This work was supported by the DFG within the Sonderforschungsbereich 634. Moreover, E. S. and W. P. S. are grateful to the German Space Agency DLR for its support within the QUANTUS collaboration with funds provided by the Federal Ministry of Economics and Technology (BMWi) under grant number DLR 50WH0837. We also thank Thomas Seligmann who induced the collaboration between the Darmstadt and the Ulm group resulting in the present article. We acknowledge fruitful and stimulating conversations with P. D\"om\"ot\"or and B. Shore. Discussions of W. P. S. with them over a time period of many years on closely related questions emerged in the scattering of a rotor from a single-slit --- a problem suggested by G. S\"u{\ss}mann and summarized by the title 'Quantum Ulm sparrow'. Moreover, W.P.S. profited from discussions with P. Brix in the early stages of this work. 
\end{acknowledgments}

\appendix
\section{\label{ssec:numerics} Numerical computation of bound states} 

\begin{figure}[bt]
\begin{center}
\includegraphics[width = 8 cm]{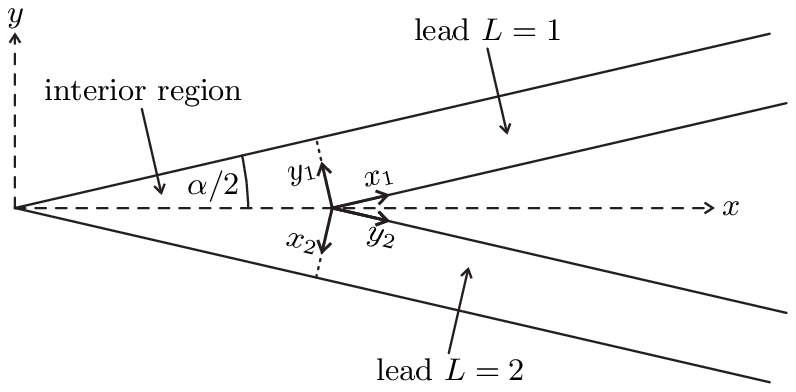}
\end{center}
\caption{\label{skizze} Definition of the two coordinate systems used in the analysis of a sharply bent waveguide. The origin of the coordinate system $(x,y)$ coincides with the outer corner and the $x$-axis with the symmetry line of the waveguide. Moreover, we identify two asymptotic regions defined by the leads and one interior region around the inner and the outer corner of the waveguide which suggests to introduce two coordinate systems $(x_1,y_1)$ and $(x_2,y_2)$ defined in the leads $L=1$ and $L=2$, respectively.}
\end{figure}

The Schr{\"o}dinger equation corresponds to that of a free particle with Dirichlet boundary conditions along the boundary $\partial\Omega$,
\be \label{eq:1} (\Delta_{x,y} + k^2) \phi(x, y) = 0, \qquad \phi(x,y)|_{\partial \Omega} = 0 \, , \ee
where $k$ is the wave number related to the energy $E$ in \refeq{eq1} as $E = k^2$. The waveguide consists of two leads 1 and 2 with parallel walls and an interior region, as depicted in Fig.~\ref{skizze}. 

The coordinate system $(x,y)$ is chosen such that its origin is located at the outer corner, and the $x$-axis coincides with the symmetry axis of the bent waveguide. In a scattering experiment waves are sent into the interior region via one lead and exit either via the same or via the other one. The scattering matrix describing this process is determined with the help of the scattering formalism applied in Refs.~\cite{Weisshaar1989,Weisshaar1991,Doron1991, Dietz1992, Schwieters1996}. With that method, the wave functions $\phi(x,y)$ solving \refeq{eq:1} are obtained by writing down general solutions valid in the leads and in the interior region, respectively, and joining them smoothly along the common chord.

The solutions $\phi(x,y)$ of \refeq{eq:1} are either symmetric or antisymmetric with respect to the $x$-axis, that is, either $\phi(x,y)=\phi(x,-y)$ or $\phi(x,y)=-\phi(x,-y)$. Hence, they vanish in the antisymmetric case along the $x$-axis, $\phi(x,0)=0$. Thus $\phi(x,y)$ obeys the Dirichlet boundary condition not only along the walls of the waveguide but also along the chord $y=0$. This fact defines a further wall which divides the bent waveguide into two straight ones with, respectively, only one open end. Consequently, waves entering one of its leads are eventually reflected specularly at the end wall and then exit again, i.e., all antisymmetric states are unbound. Since our aim is the computation of the wave numbers and the wave functions of the bound states, which have been shown to exist below the cut-off frequency of the first propagating mode, we consider only the symmetric solutions of \refeq{eq:1}. 

We introduce in each lead a coordinate system with origin at the inner corner of the bent waveguide and one axis along the inner wall of the respective lead and the other one perpendicular to it (see \reffig{skizze}). Let $L=1$ denote the lead with $y>0$ in \reffig{skizze} and $L=2$ that with $y<0$, $W$ the width of the leads and $d=W\csc(\tilde\alpha/2)$ the distance between the outer and the inner corner of the bent waveguide. 

With respect to the coordinate system $(x,y)$ the coordinates $(x_L,y_L)$ are given as
\be
\left(\begin{array}{c}x_L\\y_L\end{array}\right)=
\left(\begin{array}{cc}
\cos\varphi_L\, &\sin\varphi_L\\
-\sin\varphi_L\, &\cos\varphi_L
\end{array}\right)
\left(\begin{array}{c}x-d\\y\end{array}\right),
\label{eq:2}
\ee
where $\varphi_{L=1}=\tilde\alpha/2$ and $\varphi_{L=2}=3\pi/2-\tilde\alpha/2$. 

We introduce the notation
\be
k_m=\sqrt{k^2-\left(\frac{m\pi}{W}\right)^2},
\label{eq:3}
\ee
where $m=1,2,\dots$ denotes the transverse mode numbers in the leads.
The wave number $k_m$ is purely imaginary for $k<\frac {m\pi}{W}$ and thus does not correspond to a propagating mode. For a given $k$ the number $M_o$ of the latter equals the integer part of $Wk/\pi$, i.e.,
\be
M_o=\left[\frac{Wk}{\pi}\right]\, .
\label{eq:4}
\ee
Below the cut-off wave number $k_c$ of the first propagating mode, i.e., for $k<k_c=\frac{\pi}{W}$, no propagating modes exist in the leads.

The solutions of Eq.~(\ref{eq:1}) in the leads are given by
\be
\begin{array}{rcl}
\Psi_{n,L}^{(1)} & = & \delta_{1,L}\theta(M_o-n)\frac{e^{-ik_nx_1}}{\sqrt{\vert k_n\vert}}\frac{\sin\left(\frac{\pi n}{W}y_1\right)}{\sqrt{W/2}} \\
 & & + \sum_{m=1}^{M_o+M_c}\frac{e^{ik_mx_1}}{\sqrt{\vert k_m\vert}}\frac{\sin\left(\frac{\pi m}{W}y_1\right)}{\sqrt{W/2}}T_{mn}^{1,L}
\end{array}
\label{eq:5}
\ee
for lead 1, i.e., $y > 0$, and
\bea
\begin{array}{rcl}
\Psi_{n,L}^{(2)} & = & \delta_{2,L}\theta(M_o-n)\frac{e^{-ik_ny_2}}{\sqrt{\vert k_n\vert}}\frac{\sin\left(\frac{\pi n}{W}x_2\right)}{\sqrt{W/2}} \\
 & & +\sum_{m=1}^{M_o+M_c}\frac{e^{ik_my_2}}{\sqrt{\vert k_m\vert}}\frac{\sin\left(\frac{\pi m}{W}x_2\right)}{\sqrt{W/2}}T_{mn}^{2,L}
\end{array}
\label{eq:6}
\eea
for lead 2, i.e., $y<0$. Here, ${L_1,\, L_2} \in \{ 1, 2 \}$ refer to the leads, and $\delta_{L_2,L_1}=0$ for $L_1\ne L_2$, $\delta_{L_2,L_1}=1$ for $L_1=L_2$ and $\theta(x)=1$ for $x>0$ and zero otherwise. 

In both equations the first term corresponds to an incoming wave with mode number $n$ propagating from the asymptotic region, i.e., the ends of the leads, into the interior region, while in the second term the sum runs over all $M_o$ outgoing propagating and $M_c$ exponentially decaying modes with a purely imaginary $k_m$. Since these evanescent modes couple only weakly to the interior region for large mode numbers $m$, their total number can be limited to some finite value $M_c$. The reduced T-matrix with $n, m \leq M_o$ yields the scattering matrix $\hat S^{L_2,L_1}$. The matrix element $S_{mn}^{L_2,L_1}$ describes the scattering from mode $n$ in lead $L_1$ via the interior region into mode $m$ of lead $L_2$. 

In the interior region, the solutions of \refeq{eq:1}, which are symmetric with respect to the $x$-axis and vanish along its outer boundary, that is, along its common boundary with the bent waveguide, can be written explicitly in polar coordinates
\be \begin{array}{rcl}
x & = & r\cos\theta \\
y & = & r\sin\theta
\end{array} \label{eq:7} \ee
as
\be
\Psi^{\rm int}_n(r,\theta)=\sum_{\mu=1}^{M_t}B_{\mu n}\cos\left(\tilde\mu\theta\right)J_{\tilde\mu}(kr),
\label{eq:8}
\ee
where we introduced the notation $\tilde\mu=\frac{\pi}{\tilde\alpha}(2\mu -1)$.
Due to the property of the Bessel functions $J_{\tilde\mu}(x)$ that they become vanishingly small for $\tilde\mu\gg x$, the infinite expansion can be truncated at a certain value $\mu =M_t$, which depends on that of $kr$. 

The matrix elements $T_{mn}^{L_2,L_1}$ entering the ansatz for $\Psi_{n,L}^{(1)}$ and $\Psi_{n,L}^{(2)}$ inside the lead region given by Eqs.~(\ref{eq:5}) and~(\ref{eq:6}) and the expansion coefficients $B_{\mu n}$ entering $\Psi^{\rm int}_n(r,\theta)$ for the interior region in \refeq{eq:8} are chosen such that the wave functions $\Psi^{\rm int}_n(r,\theta)$ and $\Psi_{n,L}^{(1)}(x_1,y_1)$, respectively, $\Psi_{n,L}^{(2)}(x_2,y_2)$ and their normal derivatives match at their common boundaries represented in \reffig{skizze} by the dotted lines and defined by  $(x_1=0, 0 \leq y_1 \leq W)$ and $(0 \leq x_2 \leq W, y_2=0)$, giving rise to the conditions 
\bea
&\Psi^{\rm int}(x_1=0,y_1)=\Psi_{n,L}^{(1)}(x_1=0,y_1)\label{eq:8a} \\
&\Psi^{\rm int}(x_2,y_2=0)=\Psi_{n,L}^{(2)}(x_2,y_2=0)\label{eq:8b}
\eea
and
\bea
& \left. \frac{\partial\Psi_{}^{\rm int}}{\partial x_1}(x_1,y_1) \right|_{x_1=0} = \left. \frac{\partial\Psi_{n,L}^{(1)}}{\partial x_1}(x_1,y_1) \right|_{x_1=0}\label{eq:8c} \\
& \left. \frac{\partial\Psi_{}^{\rm int}}{\partial y_2}(x_2,y_2) \right|_{y_2=0} = \left. \frac{\partial\Psi_{n,L}^{(2)}}{\partial y_2}(x_2,y_2) \right|_{y_2=0} \, .
\label{eq:8d}
\eea
Here, $\Psi^{\rm int}(x_L,y_L)$ is expressed in terms of the coordinates of lead $L$. To compute the partial derivatives of $\Psi^{\rm int}(x,y)$ defined in Eq.~(\ref{eq:8}) with respect to these coordinates we use the relations
\be
\begin{array}{rcl}
\frac{\partial}{\partial x_1} & = & \cos(\tilde\alpha/2 -\theta)\frac{\partial}{\partial r}+\frac{\sin(\tilde\alpha/2 -\theta)}{r}\frac{\partial}{\partial\theta} \, , \\
\frac{\partial}{\partial y_2} & = & \cos(\tilde\alpha/2 +\theta)\frac{\partial}{\partial r}-\frac{\sin(\tilde\alpha/2 +\theta)}{r}\frac{\partial}{\partial\theta},
\end{array}
\ee
with the polar coordinates $r,\, \theta$ defined in \refeq{eq:7}.

Multiplying both sides of \refeq{eq:8c} with $\sin\left(\frac{\pi m}{W}y_1\right)/\sqrt{W/2}$ and of Eq.~(\ref{eq:8d}) with $\sin\left(\frac{\pi m}{W}x_2\right)/\sqrt{W/2}$ and integrating them along the common chord leaves exactly one matrix element on the right hand side of the respective equation [see Eqs.~(\ref{eq:5}) and (\ref{eq:6})]. We introduce the $(M_o+M_c)\times M_t$ dimensional matrix $\hat A$ with elements
\be
\begin{array}{rcl}
A_{n\mu} & = & \frac{\sqrt{\vert k_n\vert}}{ik_n}\int_0^{\tilde\alpha/2}{\rm d}\theta\frac{W\cot(\tilde\alpha/2)}{\cos^2(\tilde\alpha/2 -\theta)}
\frac{\sin\left(\frac{\pi n}{W}\xi\right)}{\sqrt{W/2}} \\
 & & \times \Big[ \cos(\tilde\mu\theta)\cos(\tilde\alpha/2-\theta)k \left. \frac{\partial J_{\tilde\mu}(x)}{\partial x} \right|_{x=k\rho}   \\
 & & -\sin(\tilde\mu\theta)\sin(\tilde\alpha/2 -\theta)\frac{\tilde\mu}{\rho}J_{\tilde\mu}(k\rho) \Big] \, , 
\end{array}
\ee
where
\be
\xi = -W\cot(\tilde\alpha/2)\tan(\tilde\alpha/2 -\theta) + W \, , 
\ee
with
\be
\rho = \frac{W\cot(\tilde\alpha/2)}{\cos(\tilde\alpha/2 -\theta)} \, , \label{eq:14}
\ee
for $n=1,\dots , M_o+M_c$ and $\mu =1,\dots , M_t$. 

Then we obtain for ${L_1,L_2} \in \{ 1, 2 \}$  the matrix equations
\be
-\II\delta_{L_2,L_1}+\hat T^{L_2,L_1}=\hat A\cdot\hat B.
\label{eq:15}
\ee
They relate the matrix elements $T_{mn}^{L_2,L_1}$ of $\hat T^{L_2,L_1}$ to the matrix elements $B_{\mu n}$ of $\hat B$. The values of the latter are fixed by a further set of matrix equations. To derive it, we introduce the arc length $s$, which varies along the perimeter of the interior region, and expand the wave function $\Psi^{\rm int}(x,y)$ with $x,\, y\in \partial\Omega$ in a Fourier series in $s$ (cf. Ref.~\cite{Schwieters1996}). Here, $s=y_1$ ($s=-x_2$) along the common boundary with the lead $L=1$ ($L=2$), and $s=r$ along that with the wall of the bent waveguide at $\theta=\pm\tilde\alpha/2$ in terms of the polar coordinates defined in \refeq{eq:7}. Using the fact that $\Psi^{\rm int}_n(r,\theta)$ vanishes along the outer walls, the Fourier coefficients associated with $s$ read
\be
D_{\lambda\mu}=\int_0^{\tilde\alpha/2}{\rm d}\theta\frac{W\cot(\tilde\alpha /2)}{\cos^2(\tilde\alpha/2 -\theta)}J_{\tilde\mu}(k\rho)\cos(\tilde\mu\theta)\cos\left(\frac{2\pi\lambda}{\Gamma}\xi\right)
\label{eq:16}
\ee
with $\lambda ,\, \mu =1,\dots ,\, M_t$. The variables $\xi$ and $\rho$ are defined in \refeq{eq:14} and $\Gamma=2W[1+\cot(\tilde\alpha/2)]$ denoting the length of the perimeter of the interior region. Defining the $M_t\times (M_o+M_c)$-dimensional matrix $\hat C$ with elements
\be
\begin{array}{rcll}
C_{\mu n} & = & \sqrt{\frac{W}{2}}\frac{1}{\sqrt{\vert k_n\vert}} & \left[\frac{1-\cos\left(\pi n+2\pi\mu\frac{W}{\Gamma}\right)}
{\pi n +2\pi\mu\frac{W}{\Gamma}}\right. \\
 & & & + \left.\frac{1-\cos\left(\pi n-2\pi\mu\frac{W}{\Gamma}\right)}{\pi n-2\pi\mu\frac{W}{\Gamma}}\right]
\end{array}
\label{eq:17}
\ee
with $n=1,\dots ,\, M_o+M_c,\, \mu =1,\dots ,\, M_t$, Eqs.~(\ref{eq:8a}) and (\ref{eq:8b}) yield the matrix equations
\be
\hat C\left(\II (\delta_{L,1}+\delta_{L,2}) +\hat T^{L,1}+\hat T^{L,2}\right)=2\hat D\cdot\hat B,
\label{eq:18}
\ee
for $L \in \{ 1, 2 \}$. 

Combining Eqs.~(\ref{eq:15}) and (\ref{eq:18}), the matrix elements of $\hat B$ are obtained as 
\be
\hat B=\left(\hat D-\hat C\hat A\right)^{-1}\hat C.
\label{eq:19}
\ee
For $k > k_c$, that is, if the leads allow for propagating modes, we obtain with \refeq{eq:14} setting $M_c = 0$ the final result for the scattering matrices
\bea
\hat S^{1,1}=\hat S^{2,2}=&\II +\hat A\left(\hat D-\hat C\hat A\right)^{-1}\hat C\\
\hat S^{1,2}=\hat S^{2,1}=&\hat A\left(\hat D-\hat C\hat A\right)^{-1}\hat C.
\label{eq:20}
\eea
The full scattering matrix
\be
\left(\begin{array}{cc}
\hat S^{1,1} &\hat S^{1,2}\\
\hat S^{2,1} &\hat S^{2,2}\end{array}\right)
\label{eq:21}
\ee
is symmetric according to Eq.~(\ref{eq:20}). Using that $\hat C$ and $\hat D$ are real and that $\hat A$ is purely imaginary, it can be shown that $\hat S$ is unitary. 

Below the cut-off frequency, that is, for $k < k_c$ the first term on the right hand side of Eqs.~(\ref{eq:5}) and (\ref{eq:6}) and on the left hand side of Eqs.~(\ref{eq:15}) and (\ref{eq:18}) vanishes, leading us to the equation
\be
\left(\hat D-\hat C\hat A\right)\hat B =0
\label{eq:22}
\ee
for $\hat B$, or equivalently
\be
\det\left(\hat D-\hat C\hat A\right) =0.
\label{eq:23}
\ee
The matrix elements of $\hat A$, $\hat C$ and $\hat D$ depend on $k$. Those values of $k$ which solve \refeq{eq:23} correspond to the wave numbers of the states, which are bound in the interior region. The associated wave functions are obtained by inserting these values into Eqs.~(\ref{eq:5}) and (\ref{eq:6}), omitting the first term in these equations, and into Eq.~(\ref{eq:8}). In these calculations we typically chose for the number of closed channels $M_c$ and the number of terms $M_t$ in the sum Eq.~(\ref{eq:8}) values between $1$ and $5$. 


\end{document}